\documentclass[aps,pra,twocolumn,superscriptaddress,amsfonts,amsmath,amssymb,notitlepage,floatfix]{revtex4-1}

\usepackage{graphicx,adjustbox}
\usepackage[caption=false]{subfig}
\usepackage{bbm}
\usepackage{pgfplots,tikz}
\usepackage{hyperref}

\usetikzlibrary{shapes,backgrounds,fit,decorations.pathreplacing,calc,external}
\makeatletter
\tikzset{circle split part fill/.style  args={#1,#2}{%
 alias=tmp@name, 
  postaction={%
    insert path={
     \pgfextra{%
     \pgfpointdiff{\pgfpointanchor{\pgf@node@name}{center}}%
                  {\pgfpointanchor{\pgf@node@name}{east}}%
     \pgfmathsetmacro\insiderad{\pgf@x}
      \fill[#1] (\pgf@node@name.base) ([xshift=-\pgflinewidth]\pgf@node@name.east) arc
                          (0:180:\insiderad-\pgflinewidth)--cycle;
      \fill[#2] (\pgf@node@name.base) ([xshift=\pgflinewidth]\pgf@node@name.west)  arc
                           (180:360:\insiderad-\pgflinewidth)--cycle;            
         }}}}}  
 \makeatother  
\newcommand{\boldnabla}{\mbox{\boldmath$\nabla$}}

\begin{document}


\title{Algorithm for the solution of the  Dirac equation on digital quantum computers}


\author{Fran\c{c}ois Fillion-Gourdeau}
\email{francois.fillion@emt.inrs.ca}
\affiliation{Universit\'{e} du Qu\'{e}bec, INRS-\'{E}nergie, Mat\'{e}riaux et T\'{e}l\'{e}communications, Varennes, Canada J3X 1S2}
\affiliation{Institute for Quantum Computing, University of Waterloo, Waterloo,
Ontario N2L 3G1, Canada}

\author{Steve MacLean}
\email{steve.maclean@emt.inrs.ca}
\affiliation{Universit\'{e} du Qu\'{e}bec, INRS-\'{E}nergie, Mat\'{e}riaux et T\'{e}l\'{e}communications, Varennes, Canada J3X 1S2}
\affiliation{Institute for Quantum Computing, University of Waterloo, Waterloo,
Ontario N2L 3G1, Canada}

\author{Raymond Laflamme}
\email{laflamme@iqc.uwaterloo.ca}
\affiliation{Institute for Quantum Computing, University of Waterloo, Waterloo,
Ontario N2L 3G1, Canada}
\affiliation{Department of Physics and Astronomy, University of Waterloo, Waterloo,
Ontario N2L 3G1, Canada}
\affiliation{Perimeter Institute for Theoretical Physics, Waterloo, Ontario N2L 2Y5, Canada}
\affiliation{Canadian Institute for Advanced Research, Toronto, Ontario M5G 1Z8, Canada}


\date{\today}

\begin{abstract}
A quantum algorithm that solves the time-dependent Dirac equation on a digital quantum computer is developed and analyzed. The time evolution is performed by an operator splitting decomposition technique that allows for a mapping of the Dirac operator to a quantum walk supplemented by unitary rotation steps in spinor space. Every step of the splitting method is decomposed into sets of quantum gates. It is demonstrated that the algorithm has an exponential speedup over the implementation of the same numerical scheme on a classical computer, as long as certain conditions are satisfied. Finally, an explicit decomposition of this algorithm into elementary gates from a universal set is carried out to determine the resource requirements. It is shown that a proof-of-principle calculation may be possible with actual quantum technologies.
\end{abstract}


\maketitle

\section{Introduction}
\label{sec:intro}

Quantum computing, a new paradigm in computer science, has received a lot of interests in the last few decades because it promises a significant improvement of our computational capabilities. It is now well established that certain operations could be performed in polynomial time on quantum computers, instead of exponential time for classical ones. The prospected performance of quantum computers is fully demonstrated in Shor's algorithm for the factorization of large integers \cite{shor1998}, which requires a logarithmic number of quantum gates. The success of this quintessential method has motivated the development of many other quantum algorithms and of course, has made such devices very attractive for a number of applications such as quantum simulations, cryptography and many others \cite{0034-4885-61-2-002,nielsen2010quantum}.  

The main topic of this article, which aims at finding a quantum algorithm to simulate single particle relativistic quantum mechanics governed by the Dirac equation, relates to the efficient simulation of physical quantum systems. This is one of the most important applications of quantum computing  \cite{feynman1982simulating,SLoyd,RevModPhys.86.153}. 

Quantum simulation stems from the formal analogies existing between certain Hamiltonians describing different physical objects. Exploiting this analogy, a given quantum system (the quantum computer) can be employed and programmed to emulate another one if there exists a mapping between them. In some particular cases, such a mapping can be constructed and a faithful representation can be obtained easily. For instance, when two systems are described by the same Hamiltonian with different value of physical parameters, the correspondence is direct. This technique has been employed to simulate the Dirac equation and relativistic quantum mechanics using trapped ions \cite{gerritsma2010quantum,PhysRevLett.98.253005} and optical devices \cite{PhysRevLett.105.143902,PhysRevLett.109.023602} in regimes which are usually not accessible directly. This is an example of an analog quantum simulation, which is a powerful technique but lacks some versatility because it is limited to one class of system.  

Conversely, a digital quantum computer (DQC) is made of a number of qubits. The qubits are two-state quantum entities that can serve as registers to store data. For a large class of Hamiltonians, it was demonstrated that a DQC is a universal simulator \cite{SLoyd} and thus, can be used to simulate any local quantum systems. In this case however, the mapping can be intricate because the Hamiltonian of the physical system under study can differ significantly from the one for qubits. Moreover, the qubits are discrete objects, which usually precludes the direct analogy described above. In this sense, DQCs are similar to classical computers because they operate via discrete operations and require a discrete representation of the object (here, the wave function) under investigation. For the simulation of quantum systems, this discretization process is non-unique and each discretization leads to a different numerical schemes with varying precision. The main goal of this article is to present a numerical scheme that solves the discretized Time-Dependent Dirac equation (TDDE) and which can be mapped to a $n$-qubits DQC. 

Because it describes relativistic spin-1/2 particles, the TDDE equation is important for many applications in atomic physics, heavy ion collisions, laser physics, condensed matter physics and astrophysics. In particular, its solution is required to obtain the leading order result of quantum field theory in the strong field approximation \cite{Greiner:1985}. In this work, however, this second quantized problem is not considered explicitly. Rather, the quantum algorithm is dedicated to the solution of the ``classical'' (non-second-quantized) Dirac equation. 

Solving the TDDE numerically is a challenging problem because it is a multi-dimensional hyperbolic partial differential equation with a source term. On classical computers, several numerical methods have been developed to solve this equation. A large number of approaches are based on the combined use of the split-operator and spectral schemes \cite{PhysRevA.59.604,Mocken2008868,Mocken2004558,PhysRevA.53.1605,Bauke2011,Huang2005761,Bao2004663,
Xu2013131}. Galerkin methods relying on basis function expansions and Fourier mapped methods can be found in Refs. \cite{PhysRevA.78.062711,PhysRevA.86.052705,FillionGourdeau2016122} while ``real space'' methods have been derived using finite element schemes \cite{Muller1998245,PhysRevLett.54.669} and finite difference schemes (both explicit \cite{0022-3700-16-11-017} and implicit \cite{PhysRevA.40.5548,PhysRevA.79.043418,PhysRevC.71.024904}). A leapfrog scheme on a staggered grid has also been considered \cite{Hammer201440,Hammer201450}. 

Recently, a simple split-operator scheme in real space, having connections with the Quantum Lattice Boltzmann method \cite{Succi1993327,Succi15032002,PhysRevE.75.066704,DELLAR}, was proposed and extended to second order of convergence \cite{Lorin2011190,FillionGourdeau20121403}. It was suggested that this algorithm could be efficiently implemented on a quantum computer owing to its distinctive properties \cite{PhysRevLett.111.160602}. In particular, the time evolution of the wave function proceeds by a sequence of unitary operations where rotations in spinor space are followed by space shifts. This structure is actually reminiscent of quantum walks \cite{epj2015.fillion}, the quantum analogues to classical random walks, where the ``walker'' is described by probability amplitudes \cite{QW1}. This is not a surprise because Dirac-like equations can be derived from general quantum walks by studying their continuum limit \cite{QW2,QW3,1751-8121-47-46-465302,PhysRevA.89.062109}. In this article, these properties are exploited to give an explicit quantum algorithm that solves the time-dependent Dirac equation. It is demonstrated that the time evolution can be made efficient (the number of operations required is $\mathrm{poly}(n)$, where $n$ is the number of qubits) for a certain class of external electromagnetic potentials and if certain conditions for the initialization are satisfied. A classical computer would require $O(2^{n})$ operations, making the quantum implementation exponentially faster than the classical one. 


This article is separated as follows. In Section \ref{sec:dirac_split}, the time-dependent Dirac equation is presented along with a numerical scheme based on operator splitting that allows for a numerical computation of the spinor wave function time evolution. In Section \ref{sec:wf}, the mapping of the spinor wave function on quantum registers is developed. Section \ref{sec:num_quantum_walk} is devoted to the mapping of operations obtained in the splitting operator scheme and well-known qubit gates. It is also shown that the Dirac equation evolution is a quantum walk supplemented by rotations in spinor space. In Section \ref{sec:comp_ana}, the efficiency of the algorithm is discussed and compared to the classical implementation. Section \ref{sec:init_reg} briefly mentions performance issues with the initialization of the quantum register. Section \ref{sec:ress_req} contains an explicit resource analysis where the quantum gates are decomposed into elementary gates. The conclusion is found in Section \ref{sec:conclusion} while many details on higher order split operator schemes and the initialization of the quantum register are in appendices.

\section{The Dirac equation and operator splitting}
\label{sec:dirac_split}

This section gives a review of the operator splitting method applied to the Dirac equation. More details and numerical examples can be found in Refs. \cite{Succi1993327,Succi15032002,PhysRevE.75.066704,DELLAR,Lorin2011190,FillionGourdeau20121403,PhysRevLett.111.160602}. 

The Dirac equation gives a quantum relativistic description of fermions and is the relativistic extension of the Schr\"{o}dinger equation to spin-$1/2$ particles. These particles are ubiquitous in nature and therefore, the Dirac equation has applications in many fields of physics. In this work, the focus is on the relativistic dynamics of a single electron of mass $m$ coupled to an external classical electromagnetic field characterized by its electromagnetic potential. The single particle time-dependent Dirac equation is given by \cite{Itzykson:1980rh}
\begin{align}
i\partial_t \psi(t,\mathbf{x}) = \hat{H} \psi(t,\mathbf{x}),
\label{eq:dirac_eq}
\end{align}
where 
\begin{align}
\psi(t,\mathbf{x}) = 
\begin{bmatrix}
\psi_{1}(t,\mathbf{x}) \\
\psi_{2}(t,\mathbf{x}) \\
\psi_{3}(t,\mathbf{x}) \\
\psi_{4}(t,\mathbf{x}) \\
\end{bmatrix} = 
\begin{bmatrix}
\phi(t,\mathbf{x}) \\
\chi(t,\mathbf{x}) 
\end{bmatrix},
\end{align}
is the time and coordinate dependent four-spinor, where $\phi_{1,2} = \psi_{1,2}$ are the large components and $\chi_{1,2} = \psi_{3,4}$ are the small components. The operator $\hat{H}$ is the Hamiltonian given by
\begin{align}
\hat{H}  =  \boldsymbol{\alpha} \cdot \left[  c\hat{\mathbf{p}} - e\mathbf{A}(t) \right] + \beta m c^{2} + e\mathbb{I}_{4}V(\mathbf{x},t) ,
\label{eq:hamiltonian}
\end{align}
where $e$ is the electric charge (obeying $e=-|e|$ for an electron) and the momentum operator is $\hat{\mathbf{p}} = -i \boldnabla$. Here, $\mathbf{A}(t)$ is the electromagnetic vector potential while $V(\mathbf{x},t) = A_{0}(\mathbf{x},t)$ is the scalar potential. The vector potential represents physically a time-dependent homogeneous electric field and thus, it is translation invariant and do not depend on space $\mathbf{x}$. The scalar potential, on the other hand, has a space dependence and can represent either a static (such as a Coulomb potential) or dynamic field. In this configuration, there is no magnetic field, the latter being given by $\mathbf{B} = \nabla \times \mathbf{A}$. Performance issues may arise when the magnetic field is included in the quantum algorithm, as discussed in more details in subsequent sections and in Appendix \ref{sec:magnetic}.

Finally, $\mathbb{I}_{4}$ is the 4 by 4 unit matrix and $\beta, \boldsymbol{\alpha}=(\alpha_a)_{a=x,y,z}$ are the Dirac matrices. In all calculations, the Dirac representation is used where
\begin{align}
\alpha_{a} = 
\begin{bmatrix}
	0 & \sigma_{a} \\
	\sigma_{a} & 0 
\end{bmatrix}
 \; \; , \; \;
\beta = 
\begin{bmatrix}
	\mathbb{I}_{2} & 0 \\
	0 & -\mathbb{I}_{2} 
\end{bmatrix} .
\label{eq:dirac_mat}
\end{align}
The $\sigma_{a}$ are the usual $2 \times 2$ Pauli matrices defined as
\begin{align}
\sigma_{x} = 
\begin{bmatrix}
0 & 1 \\ 1 & 0  
\end{bmatrix}
\;\; \mbox{,} \;\;
\sigma_{y} = 
\begin{bmatrix}
0 & -i \\ i & 0 
\end{bmatrix}
\;\; \mbox{and} \;\;
\sigma_{z} = 
\begin{bmatrix}
1 & 0 \\ 0 & -1 
\end{bmatrix},
\end{align}
while $\mathbb{I}_{2}$ is the 2 by 2 unit matrix.

\subsection{Time discretization}

The starting point of the general operator splitting theory is the formal solution of the Dirac equation given by
\begin{align}
\psi(t_{n+1}) &= T \exp \left[ -i \int_{t_{n}}^{t_{n+1}} \hat{H}(t) dt \right] \psi(t_{n}), \\
\label{eq:suzuki_time}
&= e^{-i\Delta t (H(t_{n}) + \mathcal{T})} \psi(t_{n}),
\end{align}
where $T$ is the time-ordering operator, $\Delta t = t_{n+1}-t_{n}$ is the time step and $\mathcal{T} = i \overleftarrow{\partial_{t_{n}}}$ is the ``left'' time-shifting operator. The second form of the solution in Eq. \eqref{eq:suzuki_time} was obtained in Ref. \cite{suzuki1993general} and constitutes a convenient starting point for deriving approximation schemes. Then, the operator splitting method consists in decomposing the Hamiltonian as $\hat{H}(t) = \sum_{j=1}^{N_{\mathrm{op}}}\hat{H}_{j}(t)$ (here, $N_{\mathrm{op}} \in \mathbb{N}^{+}$ is the number of operators) and in approximating the evolution operator in Eq. \eqref{eq:suzuki_time} by a sequence of exponentials in the form:
\begin{align}
\label{eq:approx_time}
\psi(t_{n+1})
&=  \prod_{k=1}^{N_{\rm seq}}\left[e^{-is_{0}^{(k)}\Delta t \mathcal{T}} \prod_{j=1}^{N_{\mathrm{op}}}e^{-is_{j}^{(k)}\Delta t \hat{H}_{j}(t_{n})} \right] \psi(t_{n}) \nonumber \\
& +O(\Delta t^{q}),
\end{align}   
where the coefficients $N_{\rm seq}\in \mathbb{N}^{+}$ and $s_{j}^{(k)} \in \mathbb{R}$ are chosen in order to get an approximation with a given order of accuracy $q\in \mathbb{N}^{+}$.  When some pairs of Hamiltonian in $(\hat{H}_{i})_{i=1,\cdots,N_{\mathrm{op}}}$ do not commute, the splitting induces a numerical error $O(\Delta t^{q})$, where the value of $q$ can be improved to arbitrary order \cite{suzuki1993general}. 

Such a decomposition is useful when all expressions of the form $\left. e^{itH_{j}}\right|_{j=1,\cdots,N_{\mathrm{op}}}$ can be evaluated explicitly. In principle, any decomposition can be utilized but some are particularly more convenient than others. In this work, and for reasons that will become clear later, the following decomposition is used \cite{PhysRevLett.111.160602,FillionGourdeau20121403,Lorin2011190}:
\begin{align}
\label{eq:H1}
\hat{H}_{1}  &= \hat{H}_{x} =  -ic \alpha_{x} \partial_{x} , \\
\hat{H}_{2}  &= \hat{H}_{y} =-ic \alpha_{y} \partial_{y} , \\
\hat{H}_{3}  &= \hat{H}_{z} = -ic \alpha_{z} \partial_{z} , \\
\hat{H}_{4}  &= \hat{H}_{m} =  \beta m c^{2} , \\
\hat{H}_{5}  &= \hat{H}_{V}(t) =  e\mathbb{I}_{4}V(\mathbf{x},t) \\ 
\label{eq:H5}
\hat{H}_{6}  &= \hat{H}_{\mathbf{A}}(t) = -e\boldsymbol{\alpha} \cdot    \mathbf{A}(t).
\end{align}
This corresponds to an Alternate Direction Iteration (ADI) technique whereby each direction is treated independently. Then, the following scheme with a second order accuracy can be obtained:

\begin{align}
\label{eq:first_order_scheme}
\psi(t_{n+1})
 &=  e^{-i\Delta t  \mathcal{T}}e^{-i\Delta t \hat{H}_{\mathbf{A}}(t_{n}) }e^{-i\Delta t \hat{H}_{V}(t_{n}) }e^{-i\Delta t \hat{H}_{m} } \nonumber \\
 & \times e^{-i\Delta t \hat{H}_{z} } e^{-i\Delta t \hat{H}_{y} }e^{-i\Delta t \hat{H}_{x} }  \psi(t_{n}) +O(\Delta t^{2}) ,\nonumber \\
 &= Q_{\mathbf{A}}(t_{n},\Delta t) Q_{V}(t_{n},\Delta t)Q_{m}(\Delta t) \nonumber \\
 & \times Q_{z}(\Delta t) Q_{y}(\Delta t) Q_{x}(\Delta t) \psi(t_{n}) +O(\Delta t^{2}),
\end{align}
where
\begin{align}
\left. Q_{a}(\Delta t)\right|_{a=x,y,z} &:= e^{-c\Delta t \alpha_{a}\partial_{a}}, \\
Q_{m}(\Delta t)&:=e^{-i\Delta t \beta m c^{2} }, \\
Q_{V}(t,\Delta t)&:=e^{-i\Delta t( e\mathbb{I}_{4}V(\mathbf{x},t))} ,\\
Q_{\mathbf{A}}(t,\Delta t)&:=e^{ie\Delta t\boldsymbol{\alpha} \cdot    \mathbf{A}(t)} .
\end{align}
This scheme can be improved to third and higher order accuracy by making use of a symmetric decomposition \cite{suzuki1993general,FillionGourdeau20121403,Lorin2011190}. The results for higher order are given in Appendix \ref{sec:high_order}.
This decomposition is very convenient because the effect of each operator $Q_{i}$ can be obtained exactly. For $a=x,y,z$, we define the following unitary operators: 
\begin{align}
\label{eq:rot_operator_spin}
S_{a} := \frac{1}{\sqrt{2}} (\beta + \alpha_{a}).
\end{align}  
These operators transform the Dirac matrices to a Majorana-like representation, where the 
matrix $\tilde{\alpha}_{a} =S_{a}^{\dagger}\alpha_{a}S_{a}= \beta$ is diagonal, with eigenvalues $\pm 1$. By expanding the exponential in $Q_{a}$ and introducing unit matrices in the form of $S_{a}S^{\dagger}_{a} = \mathbb{I}$, we get 
\begin{align}
Q_{a} = S_{a} T_{a} S^{\dagger}_{a},
\end{align} 
where 
\begin{align}
\label{eq:trans_op}
T_a(\Delta t)=e^{-c\Delta t \beta \partial_{a}},
\end{align}
is a translation operator along the direction $a$. The latter shifts the $\phi$ and $\chi$ spinor components by $\mp c\Delta t$, respectively. Using this result, the time evolution of the wave function is written as a sequence of unitary operators. For the first order splitting, it yields
\begin{align}
\label{eq:QLB_exp}
\psi (t_{n+1}) &= Q_{\mathbf{A}}(t_{n},\Delta t)Q_{V}(t_{n},\Delta t)Q_{m}(\Delta t)\left[S_{z}T_{z}(\Delta t)S_{z}^{-1}\right]  \nonumber \\
& \times \left[S_{y}T_{y}(\Delta t)S_{y}^{-1}\right]\left[S_{x}T_{x}(\Delta t)S_{x}^{-1}\right] \psi (t_{n}) .
\end{align}
Eqs. \eqref{eq:QLB_exp} is the most important result of this section, giving an approximation of the time evolution operator valid for small $\Delta t$. Again, this can be generalized to higher order schemes, as shown in Appendix \ref{sec:high_order}. This completes the discussion of the time discretization.

\subsection{Space discretization}

In the last section, the time discretization of the wave function was described and the time evolution was given as a sequence of unitary operations. However, to store the values of the wave function on a classical or quantum computer, the space also needs to be discretized. To be consistent with the time discretization, it is convenient to use $P_{0}$ type elements where the value of the wave function is constant within  each volume \cite{Lorin2011190}. Therefore, the space domain is discretized in cubic elements with edges of length $\ell = \Delta x = \Delta y = \Delta z$. The projection of the wave function on this grid, the discretized wave function $\psi_{\ell}$, can then be written as a tensor product of basis functions expressed as
\begin{align}
\label{eq:discr_psi}
 \psi_{\ell}(t,\mathbf{x}) &= \sum_{i=1}^{N_{x}} \sum_{j=1}^{N_{y}} \sum_{k=1}^{N_{z}} \mathbf{1}_{i}(x) \mathbf{1}_{j}(y) \mathbf{1}_{k}(z) \psi(t,\tilde{\mathbf{x}}_{i,j,k}) ,
\end{align}
where $N_{x},N_{y},N_{z}$ is the total number of intervals in each direction and $\psi_{\ell}$ is the discretized wave-function. The basis functions $\mathbf{1}_{i}(x), \mathbf{1}_{j}(y), \mathbf{1}_{k}(z)$ have a value of 1 in the  $i,j,k$ interval, respectively, and a value of zero outside. Finally,  $\tilde{\mathbf{x}}_{i,j,k}$ is the vector pointing to the centroid of each volume element. It is defined as 
\begin{align}
\tilde{\mathbf{x}}_{i,j,k} &= \biggl(x_{\rm min} + (i+\frac{1}{2})\ell, y_{\rm min} + (j+\frac{1}{2})\ell, \nonumber \\
& \;\;\;\;\;\;\; z_{\rm min} + (k+\frac{1}{2})\ell\biggr) 
\end{align}
where $x_{\rm min},y_{\rm min},z_{\rm min}$ are the lower domain boundary coordinates. 

The normalization condition then becomes
\begin{align}
\int d^{3}\mathbf{x} \psi^{\dagger}_{\ell}(t,\mathbf{x})\psi_{\ell}(t,\mathbf{x}) =1 ,
\end{align}
which is written as
\begin{align}
\label{eq:norm_wf}
 \ell^{3} \sum_{i=1}^{N_{x}} \sum_{j=1}^{N_{y}} \sum_{k=1}^{N_{z}}  \psi^{\dagger}(t,\tilde{\mathbf{x}}_{i,j,k}) \psi(t,\tilde{\mathbf{x}}_{i,j,k}) =1,
\end{align}
once it is discretized.
Thus, the amplitudes should obey $|\psi(t,\bar{\mathbf{x}}_{i,j,k})|\leq \frac{1}{\ell^{\frac{3}{2}}}$ and therefore, can be mapped easily on the finite interval $[0,1]$. This feature will be important in the next section where the mapping of the wave function on qubits is discussed. 

To keep the exactness of each step in the splitting, there is another important condition that needs to be fulfilled: the time step should be related to the space step as
\begin{align}
\label{eq:cfl}
c\Delta t = N^{*}\ell,
\end{align}
where $N^{*} = \frac{1}{2},1,\frac{3}{2},2,...$ can be any half-integer. In practice however, one chooses the smallest value as possible to preserve the efficiency of the numerical scheme. Eq. \eqref{eq:cfl} is a Courant-Friedrichs-Lewy (CFL) condition \cite{leveque2002finite}.

The value of $N^{*}$ also modifies the dispersion relation of the numerical scheme.
The latter can be evaluated for the free Dirac equation from the split operator approximation of the evolution operator, which takes the form 
\begin{align}
\psi(t_{n+1}) = U(\Delta t, -i\nabla) \psi(t_{n}),
\end{align}
where $U(\Delta t,-i\nabla)$ is an approximation of the evolution operator to some order. Here, $U$ evolves the wave function according to the free massless Dirac equation and therefore, is a product of operators $Q_{x,y,z}(\Delta t)$. Then, a Von~Neumann analysis can be performed as in Ref. \cite{Hammer201440}. The solution is assumed to be a plane wave as $\psi(t_{n})= e^{-iEt + i\mathbf{p}\cdot \mathbf{x}}$, leading to
\begin{align}
e^{-iE\Delta t} = U(\Delta t,\mathbf{p}).
\end{align}   
Diagonalizing $U(\Delta t,\mathbf{p})$, the dispersion relation is given by
\begin{align}
E\Delta t =  i \ln \left[ \lambda_{i}(\Delta t,\mathbf{p}) \right] |_{i=1,\cdots,4},
\end{align}
where $\lambda_{i}(\Delta t,\mathbf{p})  |_{i=1,\cdots,4}$ are the eigenvalues of the matrix $U(\Delta t,\mathbf{p})$. The relation between the energy and momenta is an intricate analytical formula, which is not given here for simplicity. It can be evaluated numerically however. From these results, it can be shown in 1-D ($p_{x}=p_{y}=0$) and 2-D ($p_{x}=0$) that for the value $N^{*}=\frac{1}{2}$, the scheme is free of the fermion doubling problem, i.e. zeroes of the dispersion relation in the first Brillouin zone. Then, the dispersion relation closely resembles that of the continuum, given by $E = |\mathbf{p}|$ in the free massless case. 
For other cases (in 3-D for instance), doublers may appear. As a consequence, only the low momentum mode propagation will be correctly described. The effect of the dispersion relation however can be mitigated by increasing the number of lattice points. Another way to circumvent the fermion doubling problem in 3-D is by using the reservoir method \cite{ALOUGES2002627,ALOUGES2008643}. This allows for using different values of CFL conditions, such as $N^{*}=1/4$. For this value, the fermion doubling does not arise in the second order scheme.     

The CFL condition in Eq. \eqref{eq:cfl} guarantees that the translation operators appearing in Eq. \eqref{eq:trans_op} can be treated exactly once they are discretized \cite{Lorin2011190}. In principle, other value of $N^{*}$ could be used in conjunction with other discretization scheme, but this would induce numerical diffusion which would deteriorate the solution.

Using the CFL condition, the time evolution of the discretized wave function is written as (here, the light velocity is set to $c=1$)
\begin{align}
\label{eq:QLB_exp_sp}
\psi_{\ell} (t_{n+1}) &= Q_{\mathbf{A}}\left(t_{n},\Delta t\right) Q_{V}\left(t_{n},\Delta t\right)
Q_{m}\left(\Delta t \right) \nonumber \\
& \times \left[S_{z}T_{z}(N^{*}\ell)S_{z}^{-1}\right]\left[S_{y}T_{y}(N^{*}\ell)S_{y}^{-1}\right] \nonumber \\
& \times \left[S_{x}T_{x}(N^{*}\ell)S_{x}^{-1}\right] \psi_{\ell} (t_{n}) .
\end{align}
Similar expressions can be obtained for higher order schemes (see Appendix \ref{sec:high_order}).

The outcome of applying a translation operator $T_{a}(N^{*}\ell)$ is a translation of the value of the wave function from one mesh point to the $N^{*}$'th neighbour (the $N^{*}=1/2$ corresponds to a time staggered mesh) in direction $a=x,y,z$. For the chosen CFL condition, this operation is performed exactly on the lattice: in particular, there is no approximation of the derivative.

This completes the description of the numerical scheme to solve the time-dependent Dirac equation. Of course, this can be implemented on a classical computer and it was shown that it has some important properties: in particular, it can be parallelized very efficiently \cite{FillionGourdeau20121403}. In the next sections, the implementation of this scheme on quantum computers is discussed. 

It should also be noted that the algorithm described in Eq. \eqref{eq:QLB_exp_sp} can be seen as a generalization in 3-D of the Feynman checkerboard 1-D model \cite{feynman}, which was obtained from a path integral technique. Following the argument presented in this section, the latter can be seen as a natural consequence of the operator splitting approximation and the operator decomposition in Eqs. \eqref{eq:H1} to \eqref{eq:H5}, along with the CFL condition in Eq. \eqref{eq:cfl}.

In principle, higher order schemes with a better accuracy can be found. However, for $q>3$ most of them uses irrational or complex values for $s^{(k)}_{i}$ \cite{BANDRAUK1991428,Bandrauk2006346}. Because our scheme includes streaming steps which translate the value of the wave function on the grid, such splitting cannot be used consistently to increase the order of accuracy of the Dirac solution because the translation step will not be exact and will induce numerical diffusion. Rather, a splitting where the parameters $s^{(k)}_{i}$ are rational numbers and where every $s^{(k)}_{i}$ is a multiple of the smallest one is required. Examples of higher order schemes are discussed in Appendix \ref{sec:high_order}.

\section{Quantum implementation of the split-operator scheme}
\label{sec:quantum_impl}

The quantum implementation of the split-operator method discussed in the last section requires two main features: a mapping of the wave function on a quantum register and a mapping of unitary operators on quantum gates. For the mapping of the wave function, the standard method pioneered by Zalka and Wiesner for non-relativistic quantum mechanics \cite{wiesner1996simulations,Zalka08011998,OPPROP:PR877} is employed and described in the next section. This technique has already been extended to develop quantum algorithms for the simulation of quantum systems in physics and chemistry \cite{PMID:21166541,yung2014,e12112268,kassal2008polynomial}. The mapping of unitary operators proceeds by using the analogy with quantum walks. 

\subsection{Mapping of the wave function on qubits}
\label{sec:wf}

This section is devoted to the mapping of the discretized wave function on qubits. This is required to implement the algorithm on a quantum computer: the qubits will serve as a quantum register to store the wave function, as described in \cite{Zalka08011998,OPPROP:PR877}.

The general state of $n$ qubits $|\psi_{n} \rangle$ is a vector in the Hilbert space $\mathcal{H}_{n}:=\bigotimes_{1}^{n} \mathcal{H}_{1} $ with a dimension $2^{n}$. This state can be written as
\begin{align}
|\psi_{n} \rangle 
%
&= \sum_{s_{1} = 0}^{1} \cdots \sum_{ s_{n}  = 0}^{1} \alpha_{s_{1} \cdots  s_{n}} \bigotimes_{l =1}^{n} | s_{l} \rangle,
\end{align}
where 
$\alpha_{s_{1} \cdots  s_{n}}$ are complex coefficients. 
%
%
Here, the subscripts labelling the coefficients $s_{1} \cdots  s_{n}$ are binary numbers. These coefficients are bounded by the following normalization condition:
\begin{align}
\sum_{s_{1} = 0}^{1} \cdots \sum_{ s_{n}  = 0}^{1} |\alpha_{s_{1}\cdots  s_{n}}|^{2} = 1,
\end{align}
obtained by setting the norm to $\langle \psi_{n} | \psi_{n} \rangle = 1$ and providing a probability interpretation of the wave function. 

To map the Dirac wave function on the qubits wave function, it is convenient to partition the Hilbert space in four parts as $\mathcal{H}_{n} = \mathcal{H}_{S} \otimes \mathcal{H}_{n_{x}} \otimes \mathcal{H}_{n_{y}} \otimes \mathcal{H}_{n_{z}}$. The first Hilbert space $\mathcal{H}_{S} = \mathcal{H}_{2}$ serves to label the spinor degrees of freedom. Because there are four spinor components, two qubits are reserved for this role. The other qubits will be used as quantum registers for the space dependence of the wave function. Then, the state of the quantum register is written as
\begin{align}
|\psi_{n} \rangle 
&= \sum_{s_{1},s_{2}=0}^{1}  \sum_{\{s^{(x)}\} = 0}^{1} \sum_{\{s^{(y)}\} = 0}^{1}\sum_{\{s^{(z)}\} = 0}^{1}  \nonumber \\
& \times
\alpha_{s_{1},s_{2}, \{s^{(x)}\},\{s^{(y)}\},\{s^{(z)}\}} \nonumber \\
& \times | s_{1}\rangle \otimes | s_{2}\rangle 
\bigotimes_{l_{x} =1}^{n_{x}} | s_{l_{x}} \rangle,
\bigotimes_{l_{y} =1}^{n_{y}} | s_{l_{y}} \rangle
\bigotimes_{l_{z} =1}^{n_{z}} | s_{l_{z}} \rangle,
\end{align}
where $\{s^{(a)}\} := s^{(a)}_{1} \cdots s^{(a)}_{n_{a}}$ is the set of all qubits that label the space dependent part of the wave function in the coordinate $a=x,y,z$.  
This equation can be re-written as
\begin{align}
\label{eq:coeff_regis}
|\psi_{n} \rangle = \sum_{S=1}^{4} \sum_{i=1}^{N_{x}} \sum_{j=1}^{N_{y}} \sum_{k=1}^{N_{z}} \alpha_{S,i,j,k} | S \rangle \otimes |i,j,k \rangle, 
\end{align}
where $n_{x} + n_{y} + n_{z} = n - 2 $ and $N_{x,y,z} = 2^{n_{x,y,z}}$. 
Here, we have redefined the coefficient subscripts as
\begin{align}
(i-1)_{10} &:= (s^{(x)}_{1} \cdots s^{(x)}_{n_{x}})_{2}, \\
(j-1)_{10} &:= (s^{(y)}_{1} \cdots s^{(y)}_{n_{y}})_{2},\\
(k-1)_{10} &:= (s^{(z)}_{1} \cdots s^{(z)}_{n_{z}})_{2},
\end{align}
where the notation $(b)_{n_{b}}$ stands for the number $b$ expressed in base $n_{b}$. Also, the first qubit labels whether the spinor component is a large or small component, while the second labels the component itself. Thus, we have
\begin{align}
|S\rangle := | s_{1} \rangle \otimes | s_{2}\rangle ,
\end{align}
where $s_{1} = \phi,\chi$ and $s_{2} = 1,2$.

This section is concluded with the mapping between the discretized wave function and qubits which can be written as
\begin{align}
\mathbf{1}_{i}(x) &\mapsto  |i\rangle ,\\
\mathbf{1}_{j}(y) &\mapsto  |j\rangle ,\\
\mathbf{1}_{k}(z) &\mapsto  |k\rangle ,\\
\ell^{\frac{3}{2}}\psi_{S}(t,\tilde{\mathbf{x}}_{i,j,k}) &\mapsto  \alpha_{S,i,j,k}. 
\end{align}
In other words, there are $n-2$ qubits utilized to label the space degree of freedom and to replace the basis functions. In each space dimension, there are $n_{x,y,z}$ qubits, allowing to store $N_{x,y,z}$ discretization points. Moreover, the wave function needs to be scaled by $\ell^{\frac{3}{2}}$ to have the same norm as qubits (see Eq. \eqref{eq:norm_wf}).

\subsection{Numerical scheme as a conditional quantum walk}
\label{sec:num_quantum_walk}

The numerical scheme obtained to solve the Dirac equation has three types of operator: rotation operators in spin space $S_{a}$, translation operators $T_{a}$ and mass-like local operators $Q_{m}$, $Q_{V}$ and $Q_{\mathbf{A}}$. Their mapping on a quantum computer is now discussed.

\subsubsection{Rotation operator in spinor space}

\begin{figure}
\includegraphics[scale=1.0]{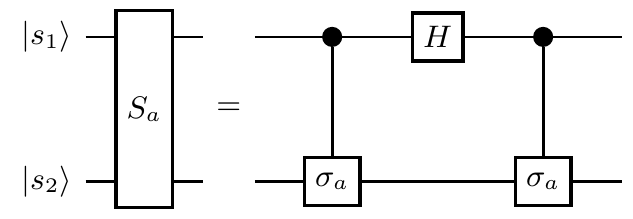}
\caption{Circuit diagram for the rotation operator in spinor space. The other qubits of the register ($|i\rangle,|j\rangle,|k\rangle$) are not modified by this transformation.}
\label{fig:rot_op}
\end{figure}

The rotation operator is given in Eq. \eqref{eq:rot_operator_spin} and is expressed in terms of Dirac matrices. Therefore, it operates in spinor space only and as a consequence, the equivalent operator in the qubit Hilbert space is different from the identity only for the spinor subspace $\mathcal{H}_{S}$. 
%
%
%
%
In this subspace and in the computational basis, the rotation matrices are given by
\begin{align}
S_{a} := \frac{1}{\sqrt{2}} 
\begin{bmatrix}
\mathbb{I}_{2} & \sigma_{a} \\
\sigma_{a} & -\mathbb{I}_{2}
\end{bmatrix}.
\end{align}
Thus, the rotation operators are represented by 2-qubits gates acting on the first two qubits. The last matrix can be decomposed as
\begin{align}
\label{eq:rot_gate_spin}
S_{a} = C(\sigma_{a})(H\otimes \mathbb{I}_{2})C(\sigma_{a}),
\end{align}
where $H$ is the Hadamard gate while $C(\sigma_{a})$ is the controlled-$\sigma_{a}$ gate. The equivalent quantum circuit is displayed in Fig. \ref{fig:rot_op}.

\subsubsection{Translation operator: quantum walk}

\begin{figure}
\includegraphics[scale=1.0]{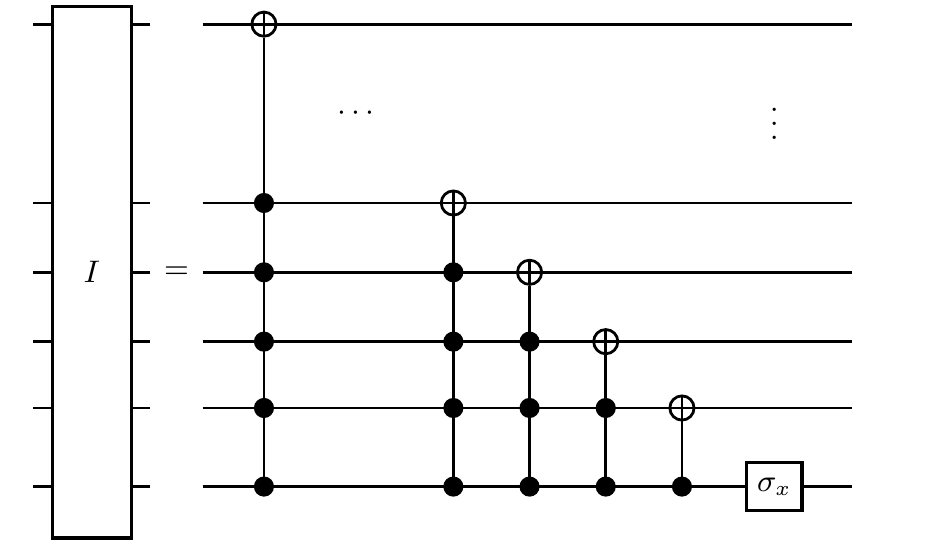} \\
\includegraphics[scale=1.0]{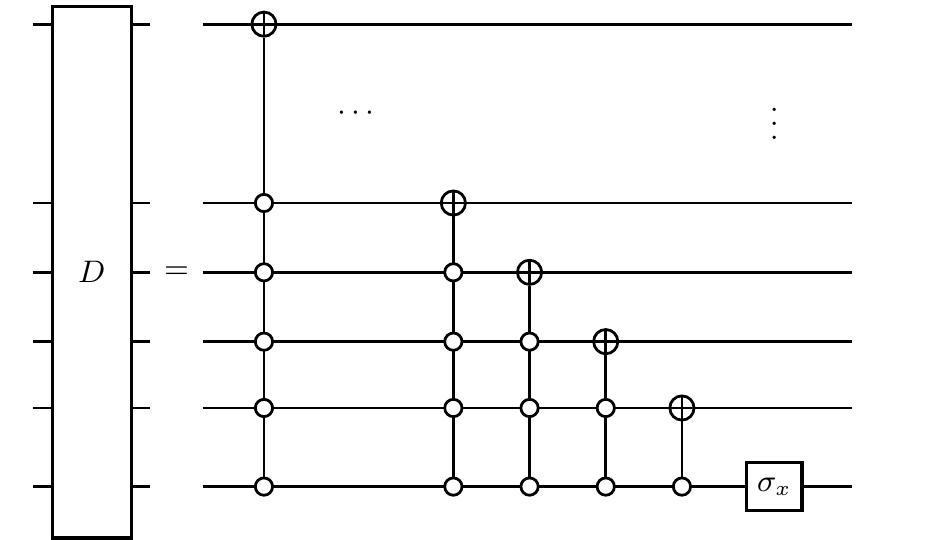}
\caption{Circuit diagram for the increment and decrement operator acting on qubits storing the data for a given dimension (the set of $n_{a}$ qubit). This implementation is borrowed from \cite{PhysRevA.79.052335}.}
\label{fig:incr}
\end{figure}

The translation operators induce a general quantum walk \cite{epj2015.fillion} because $T_{a}$ translates the small or large spinor components by $l$ step. Therefore, on the qubit Hilbert space, they can be defined as a conditional shift operator as \cite{PhysRevA.72.062317}
\begin{align}
\label{eq:incrm_x}
T_{x} |\phi \rangle \otimes | s_{2} \rangle \otimes |i,j,k \rangle &= |\phi \rangle \otimes | s_{2} \rangle \otimes |i \ominus l,j ,k \rangle ,\\
T_{x} |\chi \rangle \otimes | s_{2} \rangle \otimes |i,j ,k \rangle &= |\chi \rangle \otimes | s_{2} \rangle \otimes |i \oplus l,j,k \rangle ,\\
T_{y} |\phi \rangle \otimes | s_{2} \rangle \otimes |i,j,k \rangle &= |\phi \rangle \otimes | s_{2} \rangle \otimes |i,j \ominus l ,k \rangle ,\\
T_{y} |\chi \rangle \otimes | s_{2} \rangle \otimes |i,j ,k \rangle &= |\chi \rangle \otimes | s_{2} \rangle \otimes |i,j \oplus l,k \rangle ,\\
T_{z} |\phi \rangle \otimes | s_{2} \rangle \otimes |i,j ,k \rangle &= |\phi \rangle \otimes | s_{2} \rangle \otimes |i ,j ,k \ominus l\rangle ,\\
\label{eq:decrm_z}
T_{z} |\chi \rangle \otimes | s_{2} \rangle \otimes |i,j ,k \rangle &= |\chi \rangle \otimes | s_{2} \rangle \otimes |i,j ,k \oplus l\rangle .
\end{align}
The first qubit controls the shift because it determines which of the large or small component gets translated: the large component is shifted upward while the small component is shifted downward. These operations can be represented and decomposed efficiently into quantum gates by using controlled increment and decrement operators where the control is on the qubit $|s_{1} \rangle $, determining which of the small or large component is shifted.  The shift can be performed by using a set of controlled gates on qubits  \cite{PhysRevA.79.052335}, as displayed in Fig. \ref{fig:incr}. The full controlled shifting operation induced by the operator $T_{a}$ is depicted in Fig. \ref{fig:timestep}.  

\subsubsection{Local mass-like operators}

The evolution of the wave function by one time step requires the application of local operators $Q_{m}$, $Q_{V}$ and $Q_{\mathbf{A}}$. These operators are responsible for the mass term, the scalar potential term and the coupling of the fermion to an homogeneous electromagnetic field, respectively. 

The mass operator is simply given by
\begin{align}
Q_{m}(\Delta t) &= e^{-i\beta \Delta t mc^{2}}.
\end{align}
This operator is uniformly applied on all positions. Therefore, on the quantum register, it can be represented by
\begin{align}
\label{eq:mass_op}
Q_{m}(\Delta t) = R_{z}(2mc^{2} \Delta t) \otimes \mathbb{I}_{2} \otimes \cdots \otimes \mathbb{I}_{2}
\end{align}
where $R_{z}(\theta) := e^{-i \frac{\theta}{2}\sigma_{z}}$ is the rotation operator. The rotation is applied on the qubit $|s_{1}\rangle$, as displayed in Fig. \ref{fig:timestep}.

\begin{figure*}
\includegraphics[scale=1.0]{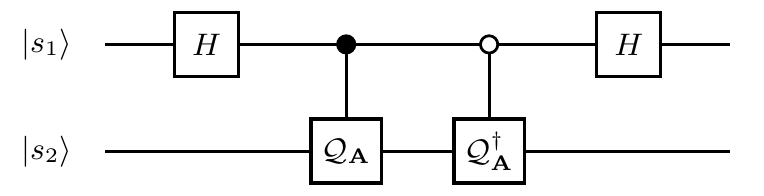} \\
\includegraphics[scale=1.0]{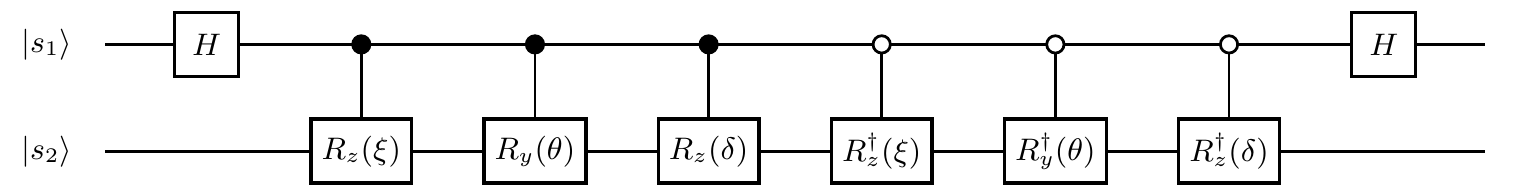}
\caption{Circuit diagram that implements the local operator $Q_{\mathbf{A}}$. The bottom picture displays explicitly the decomposition of the matrices $\mathcal{Q}_{\mathbf{A}}$ and $\mathcal{Q}^{\dagger}_{\mathbf{A}}$, appearing in the top picture.}
\label{fig:local_elec}
\end{figure*}

The second local operator, responsible for the electromagnetic potential, requires slightly more work. First, a transformation has to be applied to change the representation of Dirac matrices:
\begin{align}
Q_{\mathbf{A}}(t,\Delta t) &= (H \otimes \mathbb{I}_{2})\tilde{Q}_{\mathbf{A}}(t,\Delta t)(H \otimes \mathbb{I}_{2}),
\end{align}
where $H$ is again the Hadamard matrix. This transformation allows one to write the Dirac matrices as
\begin{align}
\alpha_{a} = (H \otimes \mathbb{I}_{2})\tilde{\alpha}_{a}(H \otimes \mathbb{I}_{2}),
\end{align}
where the Dirac matrices $(\tilde{\alpha}_{a})_{a=x,y,z}$ are now expressed in the Weyl (or chiral) representation. This representation is given by
\begin{align}
\tilde{\alpha}_{a} = 
\begin{bmatrix}
	\sigma_{a} & 0 \\
	0 & -\sigma_{a} 
\end{bmatrix}.
\end{align}
As a consequence, the local operator is also expressed in Weyl representation as
\begin{align}
\label{eq:Q_A01}
\tilde{Q}_{\mathbf{A}}(t,\Delta t)&:=e^{i\Delta t(e\tilde{\boldsymbol{\alpha}} \cdot    \mathbf{A}(t))}, \\
&= 
\begin{bmatrix}
\mathcal{Q}_{\mathbf{A}}^{\dagger} & 0\\
0& \mathcal{Q}_{\mathbf{A}} ,
\end{bmatrix},
\end{align}
where $\mathcal{Q}_{\mathbf{A}}(t,\Delta t):=e^{-ie\Delta t \boldsymbol{\sigma} \cdot    \mathbf{A}(t)}$. The operator $Q_{\mathbf{A}}$ can then be implemented as a sequence of two controlled quantum gates and Hadamard gates, as displayed in Fig. \ref{fig:local_elec}.

The last step of the quantum implementation of local operators is the decomposition of the matrix $\mathcal{Q}_{\mathbf{A}} \in SU(2)$. As noted in Ref. \cite{epj2015.fillion}, $\mathcal{Q}_{\mathbf{A}}$ is expressed in the canonical representation of the $SU(2)$ group obtained by the exponential mapping of the Lie algebra:
\begin{align}
\label{eq:V_can}
\mathcal{Q}_{\mathbf{A}}(t,\Delta t) = \mathbb{I}_{2} \cos(|\mathbf{A}(t)|\Delta t) -i \frac{\boldsymbol{\sigma}\cdot \mathbf{A}(t)}{|\mathbf{A}(t)|}  \sin(|\mathbf{A}(t)|\Delta t), \nonumber \\
\end{align}
where $|\mathbf{A}(t)| = \sqrt{A_{x}^{2}(t) + A_{y}^{2}(t) + A_{z}^{2}(t)}$.
This representation does not have a direct quantum gates decomposition. Rather, the Euler-angle parametrization is much more convenient because it can be decomposed in three rotation operators. Therefore, it is possible to write 
\begin{align}
\label{eq:V_euler}
\mathcal{Q}_{\mathbf{A}}(t,\Delta t) &= R_{z}(\delta)R_{y}(\theta)R_{z}(\xi),
\end{align}  
where $R_{a}(\theta') := e^{-i\sigma_{a}\frac{\theta'}{2}}$ is a rotation operator and $\delta,\theta,\xi \in \mathbb{R}$ are three rotation parameters. To complete the connection between the two representations, these parameters have to be linked to the three parameters characterizing the canonical representation. Using Eqs. \eqref{eq:V_can} and \eqref{eq:V_euler}, it is possible to show that \cite{epj2015.fillion}
\begin{align}
\label{eq:rot_angle_delta}
\delta &= \arctan\left[\frac{A_{z}(t)}{|\mathbf{A}(t)|} \tan \left(|\mathbf{A}(t)| \Delta t \right) \right] - \arctan\left[\frac{A_{x}(t)}{A_{y}(t)} \right],\\
\label{eq:xi}
\xi &= \arctan\left[\frac{A_{z}(t)}{|\mathbf{A}(t)|} \tan \left(|\mathbf{A}(t)| \Delta t \right) \right] + \arctan\left[\frac{A_{x}(t)}{A_{y}(t)} \right], \\
\label{eq:rot_angle_theta}
\theta &= 2\arccos \left[\frac{\cos\left(|\mathbf{A}(t)| \Delta t \right)}{\cos \left( \frac{\xi + \delta}{2} \right)} \right].
\end{align}
This gives the gate decomposition given at the bottom of Fig. \ref{fig:local_elec}. If the vector potential is space dependent to accommodate for a magnetic potential, the circuit has to be modified by adding uniformly controlled gates (see Appendix \ref{sec:magnetic}).

The last local operator $Q_{V}$ is a space-dependent phase operation, similar to the one found for the simulation of the single-particle Schr\"{o}dinger equation \cite{Zalka08011998,Strini2008}. When translation invariance is imposed and the scalar potential does not depend on space, this operator can be omitted altogether because it becomes a global phase. However, if the potential has space dependence, it has to be considered explicitly but the details of the quantum implementation depends on its functional form. The quantum evaluation of $Q_{V}$ is thus an oracle call. Because it does not mix spinor components nor change their signs and amplitudes, a general $Q_{V}$ will be decomposed using gates acting on the subspace $\mathcal{H}_{n_{x}} \otimes \mathcal{H}_{n_{y}} \otimes \mathcal{H}_{n_{z}}$, as displayed in Fig. \ref{fig:timestep}. It is not possible to implement a generic $V(\mathbf{x},t)$ because this entails the usage of diagonal unitary operators, which require an exponential number of $O(2^{n+1})$ quantum gates \cite{bullock2004asymptotically}. Nevertheless, some physically relevant potentials can be implemented efficiently. For example, polynomial potentials of a given order $k$ can be implemented in $O(n^{k})$ operations \cite{Strini2008} while an efficient gate count for the Coulomb potential can be found \cite{kassal2008polynomial}. 

When the electromagnetic field is homogeneous, it is also possible to consider a gauge in which the vector potential is $\mathbf{A}(t)=0$ while the scalar potential is $V(\mathbf{x},t) = -\mathbf{x} \cdot \mathbf{E}(t)$. In this case, the operator $Q_{\mathbf{A}}$ is not needed but the number of operations to implement the linear function in the scalar potential is $\mathrm{poly}(n_{x},n_{y},n_{z})$. Conversely, in the temporal gauge ($V(\mathbf{x},t)=0$), the vector potential is space independent and thus, the number of quantum gates is independent of the number of lattice points. As a consequence, the latter case will require less quantum operations and will be more efficient.

\subsection{Complexity analysis of the algorithm}
\label{sec:comp_ana}

\begin{figure*}
\includegraphics[scale=1.0]{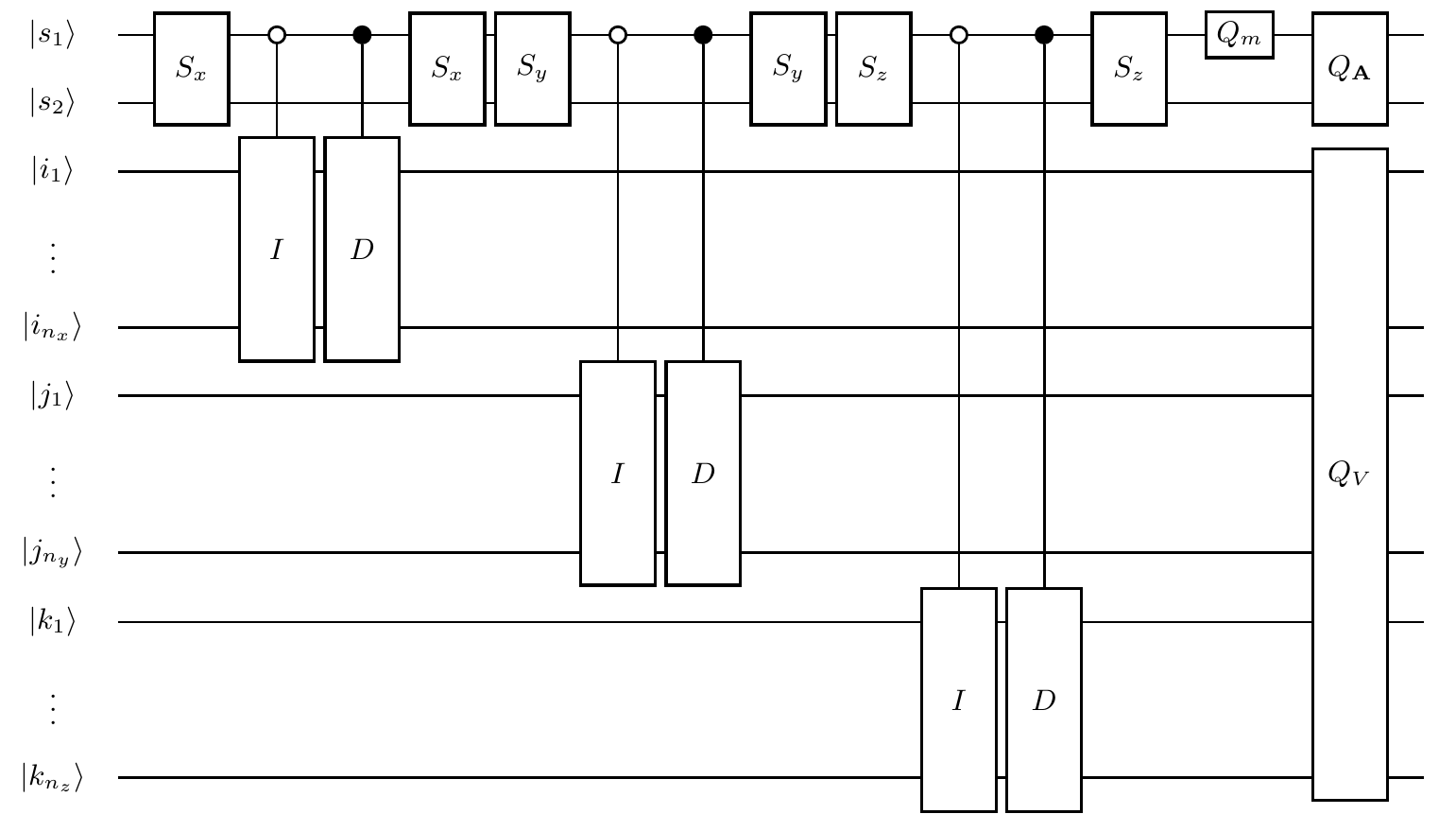}
\caption{Circuit diagram that evolves the wave function by one time step. The details of the spin rotation gates $H_{a}|_{a=x,y,z}$ are given in Fig. \ref{fig:rot_op} and Eqs. \eqref{eq:rot_gate_spin}. The increment and decrement operators $I/D$ are detailed in Fig. \ref{fig:incr} and Eqs. \eqref{eq:incrm_x} to \eqref{eq:decrm_z}. The mass operator $Q_{m}$ is given in Eq. \eqref{eq:mass_op}. Finally, the operator $Q_{\mathbf{A}}$ is displayed in Fig. \ref{fig:local_elec} and is explicitly written in Eqs.  \eqref{eq:Q_A01} to \eqref{eq:xi}. }
\label{fig:timestep}
\end{figure*}

The full time evolution for the second order accuracy scheme (the third order is a straightforward extension) is displayed in Fig. \ref{fig:timestep}. The first three groups of four gates implement the quantum walk part while the last two gates are local collision operators. As demonstrated in the following, this quantum algorithm is efficient because it scales like $\mathrm{poly}(n)$.  

The number of operations for the application of rotation operators and local operators is constant and does not augment as the number of lattice points is increased, except for the oracle $Q_{V}$. However, as discussed earlier, physically relevant potentials can be implemented efficiently using $\mathrm{poly}(n)$ number of gates. To preserve the efficiency of the algorithm, only this type of potentials is considered.

The computational complexity for the increment and decrement operators is now discussed. Classically, they can be implemented by using $N_{a}-1$ (here, $a=x,y,z$) \texttt{SWAP} operations and therefore, they have a linear classical scaling with the number of lattice points. 
 
On the other hand, the quantum implementation of the increment and decrement operators can be expressed in terms of one- and two-qubits gates taken from a universal set. It was demonstrated that a generalized \texttt{CNOT} gate, as the ones appearing in the increment and decrement operators, can be simulated by $O(n_{a})$ basic operations \cite{PhysRevA.52.3457}. However, to generate the quantum walk, one needs $n_{a}-1$ of these gates, plus one $\sigma_{x}$ gate. Therefore, the number of operations required should scale like $O(n_{a}^{2}) = \mathrm{poly}(n_{a})$, for any coordinate. Then, for $N_{a}$ lattice points in a given dimension $a$, the number of operations scales like $O(\log_{2}^{2}N_{a}) = O(n_{a}^{2})$. This corresponds to an exponential speedup of the quantum computation, in comparison to the classical case, for the increment and decrement parts. It also proves that the time evolution of the wave function displayed in Fig. \ref{fig:timestep} can be simulated with a number of gates $N_{\mathrm{gate}}$ obeying $N_{\mathrm{gate}} = \mathrm{poly}(n_{x},n_{y},n_{z})$. This is exponentially better than the classical implementation which scales like  $N_{\mathrm{op}} = \mathrm{poly}(N_{x},N_{y},N_{z})$. The asymptotic scaling behavior of the Dirac evolution algorithm will be demonstrated explicitly in Section \ref{sec:ress_req}. Of course, after $N_{t}$ time iterations, the number of quantum gates becomes $N_{\mathrm{gate}}(N) = N_{t} \mathrm{poly}(n_{x},n_{y},n_{z})$.
Then, following Refs. \cite{PhysRevA.88.022316,Ronnow420}, the quantum speedup $S_{1}(N)$ is defined as 
$S_{1}(N) = \lim_{N \rightarrow \infty} N_{\mathrm{op}}(N)/ N_{\mathrm{gate}}(N)$. Therefore, our quantum algorithm has an exponential speedup over its classical counterpart.

The previous estimates and comparisons are performed for a fixed numerical error $\epsilon$, which takes the same value in the quantum and classical implementations. In both cases, the error decreases polynomially with the number of lattice points because the time and space steps are related by the CFL condition, implying that $N_{t} = O(N)$, where $N:=N_{x}N_{y}N_{z}$ is the total number of discretization points. Then, assuming that the operators in the splitting are smooth enough and that the norm of the operator exponentials are bounded by one \cite{1751-8121-43-6-065203}, the error after $N_{t}$ iterations scales like $\epsilon = O(N_{t} \Delta t^{q}) = O(1/N_{t}^{q-1}) = O(1/N^{q-1})$. Using these results, the number of gates scales like $N_{\mathrm{gate}}(N) = N \mathrm{poly}(\log_{2}(N_{x}),\log_{2}(N_{y}),\log_{2}(N_{z}))$ while in the classical case, we have $N_{\mathrm{op}}(N) = N\mathrm{poly}(N_{x},N_{y},N_{z})$. In terms of the precision, we get (as $\epsilon \rightarrow 0$)
\begin{align}
N_{\mathrm{gate}}(\epsilon) =  \epsilon^{-\frac{1}{q-1}} \mathrm{poly}\left[ \log_{2}\left(\epsilon^{-\frac{1}{q-1}}\right)\right],
\end{align} 
for the quantum algorithm while in the classical case, one gets that
\begin{align}
N_{\mathrm{op}}(\epsilon) = \epsilon^{-\frac{1}{q-1}} \mathrm{poly}\left[ \epsilon^{-\frac{1}{q-1}}\right].
\end{align} 
Therefore, even if the CFL condition links the time and space steps, the algorithm has a strong exponential speedup, defined as $S_{2}(N) = \lim_{\epsilon \rightarrow 0} N_{\mathrm{op}}(\epsilon)/ N_{\mathrm{gate}}(\epsilon)$ \cite{PhysRevA.88.022316}.  The advantage of the quantum approach will be exhibited explicitly in Section \ref{sec:ress_req} where a gate decomposition of the algorithm will be presented.

\subsection{Initialization of the quantum register}
\label{sec:init_reg}

Before utilizing the quantum algorithm described in previous section, the quantum register has to be initialized to a physically relevant state $\psi_{S,\mathrm{init}}(\mathbf{x})$. This is performed by setting the coefficients $\alpha_{S,i,j,k}$ that encode the wave function, to some properly chosen value.
This can be a challenge because initializing general states require diagonal unitary operations. As demonstrated in Appendix \ref{sec:gen_init}, the quantum gate decomposition for the initialization of a general wave function is given in terms of uniformly controlled gates. The optimal number of gates required to carry these operations scales like $O(2^{n+1})$ \cite{PhysRevLett.91.027902,PhysRevA.71.052330,bullock2004asymptotically}, which will obliterate the performance of the quantum time-evolution algorithm. However, this technique can be useful to simulate elementary quantum systems \cite{Strini2008}. 

For many physical applications, it is enough to start the simulation with an eigenstate of some static potential instead of some general state \cite{tannor2007introduction}. In this case, the phase-estimation method can be employed and can be implemented efficiently under some conditions \cite{PhysRevLett.83.5162,kassal2008polynomial,Aspuru-Guzik1704}. This procedure allows for the determination of both the eigenvalues and eigenstates.  However, it requires many ancilla qubits to have enough energy resolution. The number of ancilla qubits can be reduced significantly by a filtering technique inspired from the Feit-Fleck method \cite{ffg.feitfleck}. The latter is described in Appendix \ref{sec:init}.

\section{Resource requirements and feasibility}
\label{sec:ress_req}

The circuit depth (number of gates) and width (the number of required ancilla qubits) for the quantum Dirac solver is now determined by using \texttt{Quipper} to perform an explicit gate decomposition \cite{green2013quipper}.  \texttt{Quipper} is a functional scalable quantum programming language capable of, among other things, translating intricate quantum algorithms and circuits into sequences of elementary gates from a given universal set. Moreover, it includes many functions to specify and manipulate quantum circuits. Therefore, it is an efficient and convenient tool for the concrete determination of quantum resource requirements of a given quantum algorithm. As a matter of fact, it has been utilized to analyze the resource requirement for some common quantum algorithms such as the  quantum linear system algorithm \cite{scherer2015resource} and others \cite{green2013quipper,siddiqui2014five}. Here, the feasibility of the implementation of the Dirac solver on actual quantum computers, for a proof-of-principle calculation, is assessed with this tool. 

In the following results, an idealized quantum computer is assumed where all the quantum operations are carried without error. In a real device, some errors could be occurring due to noise coming from the interaction with the environment. These errors can be compensated by error correcting algorithms but this demands for more resources. In this sense, the results given in the following are lower bound estimate for real calculations.  

The algorithm given in Section \ref{sec:quantum_impl} and more precisely, the part for the time evolution displayed in Fig. \ref{fig:timestep}, is coded in \texttt{Quipper}. For simplicity and because it depends on the physical system studied, a vanishing scalar potential is assumed ($V(\mathbf{x},t) = 0$). The  contribution of this oracle can be evaluated independently for specific applications. The quantum gates in the Dirac solver algorithm are decomposed into a standard universal set of gates comprising the Hadamard (\texttt{H}), the Clifford (\texttt{S}), the $\frac{\pi}{8}$-phase (\texttt{T}) and the controlled-not (\texttt{CNOT}) gates. Hereinafter, these gates will be denoted as fundamental quantum gates. These gates are then used to approximate all the logical gates in our algorithm. For rotation gates appearing in mass operators, a numerical precision is required and is set to 10 digits. Of course, a higher precision will entail a larger number of fundamental gates. The value of the vector potential is set to an arbitrary value while the time step is set to $\Delta t = 0.0001$, although the explicit value of $\Delta t$ does not have a large effect on the gate count. 

\begin{figure}
\includegraphics[width=0.45\textwidth]{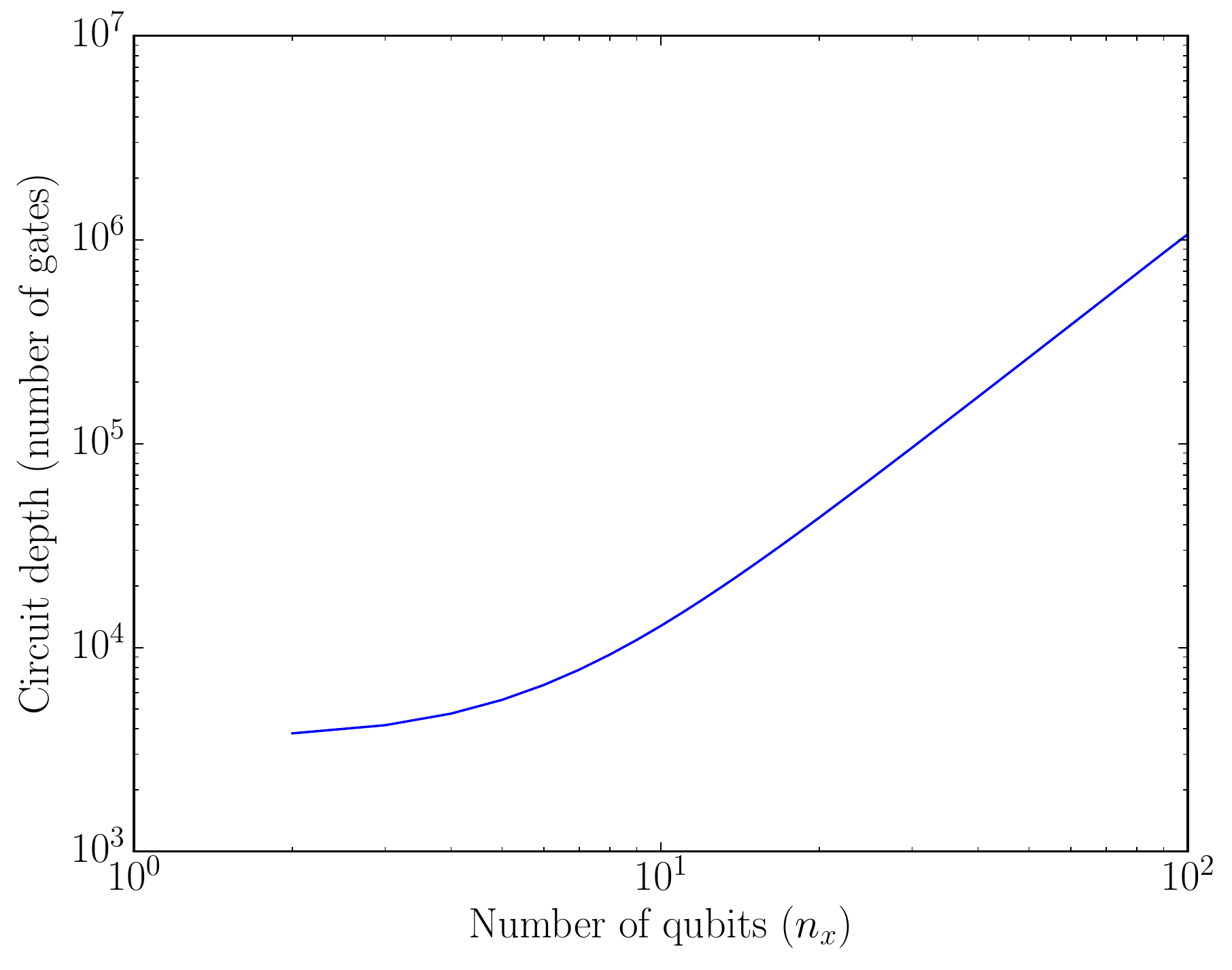}
\caption{The circuit depth (number of gates) required to evolve the wave function by one time step as a function of the number of qubits. The number of qubits corresponds to $n_{x}$. It is assumed that $n_{x} = n_{y} = n_{z}$. The circuit width required is then given by $n_{\mathrm{total}} = 4n_{x}+2$. }
\label{fig:ress_req}
\end{figure}

The results for the circuit depth as a function of the number of qubits are displayed in Fig. \ref{fig:ress_req}. The number of gates is obtained from the decomposition into the fundamental set of gates while the number of qubits displayed in the figure corresponds to $n_{x}$, i.e. the number of qubits used to store the wave function $x$-coordinates. It is assumed that $n_{x} = n_{y} = n_{z}$. Moreover, the evaluation of multi-controlled gates in the increment and decrement operators requires $n_{x}$ ancilla qubits, making for a circuit width (total number of qubits) of $n_{\mathrm{total}} = 4n_{x}+2$.

It is verified by fitting the data in Fig. \ref{fig:ress_req} with a polynomial that for a large number of qubits ($n_{x} \gtrsim 10$), the number of gates increases quadratically. This confirms the complexity analysis and the asymptotic behavior given in Section \ref{sec:comp_ana}. For $n_{x} \lesssim 10$, the dependence is close to a linear behavior. For any number of quantum qubits, there is a given number of quantum gates reserved for the local mass operators. In particular the gate $Q_{m}(\Delta t)$ requires 245 fundamental quantum gates while the $Q_{\mathbf{A}}$ necessitates 3330 fundamental quantum gates. If desired, the circuit depth for these gates could be reduced by decreasing the precision for the approximation of rotation operators.

The maximum number of qubits considered in this analysis is $n_{x} = 100$ ($n_{\mathrm{total}} = 402$), corresponding to a simulation with a lattice of size $ N_{x} = N_{y} = N_{z} = 1.27 \times 10^{30} $. This is much higher than the number of lattice points any classical computer can accommodate. Moreover, assuming that the number of classical operations is linear with the number of lattice points (see Section \ref{sec:comp_ana}), the total number of operations on a classical computer would reach $\sim 10^{90}$, which is many orders of magnitude higher than on a quantum computer, which necessitates $\sim 10^{6}$ quantum gates. State of the art classical simulations of the Dirac equation could use lattice with a size $ N_{x} = N_{y} = N_{z} = 1024$ \cite{FillionGourdeau20121403}, requiring approximately $\sim 10^{9}$ operations. In comparison, our quantum algorithm would need $n_{x}=10$ ($n_{\mathrm{total}} = 42$) and 12773 quantum gates, which is five orders of magnitude below. These comparisons clearly attest to the advantage of the quantum computer over the classical computer.

However, actual quantum devices are limited in the number of qubits and the number of gates they can apply on the quantum register. The maximum number of entangled qubits is $\sim 14$ \cite{PhysRevLett.106.130506} while the maximum number of gates (quality factor) reaches $\sim 10^{4}$ \cite{0034-4885-74-10-104401}. Combining a high quality factor along with using a large number of qubits is a challenging experimental problem. Nevertheless, digital quantum computing has been conducted on various devices. For example, it has been accomplished with a superconducting circuit using nine qubits and $\approx 1000$ quantum logic gates \cite{barends2016digitized,barends2015digital,kelly2015state}. Trapped ions have also been considered, where $\approx 100$ quantum logic gates on six qubits have been achieved \cite{Lanyon57}. Finally, circuit quantum electrodynamics has been utilized to simulate quantum interacting spin models using two qubits \cite{PhysRevX.5.021027}.

Given these limitations, we now try to find some parameters which could allow for a proof-of-principle simulation. The simplest system that can be studied is the massless ($m=0$) 1-D electron. In this case, there are no rotation operators, reducing the relative number of operation significantly for low number of qubits. The mass term can be included, adding 130 gates (using a precision of 5 digits) to the massless case.  In 1-D, two spinor components become redundant and can be discarded, also reducing the number of qubit by one \cite{epj2015.fillion}. The resource requirements for this system are given in Table \ref{table:ress_req} for many lattice sizes. Clearly, the lattice size that can be simulated on actual quantum computer is relatively small compared to classical calculations. More importantly however is the number of time steps which can be simulated, on the order of $\sim $10-25 for lattice size of 16-32 points. This is much lower than the number of time steps usually required in classical simulations, which reaches 1000-10000 time steps for much larger lattice sizes \cite{FillionGourdeau20121403}. Advancing the wave function by 10-25 time steps may be enough for a proof-of-principle calculation using present day technologies, but this is not sufficient for conducting physically relevant calculations. It also demonstrates that quantum computers are still far from outperforming classical calculations.

\begin{table}[h]
\caption{Minimal resource requirements for simulating the 1-D massless Dirac equation on an actual quantum computer device. With the mass term, the circuit depth increases by 130 gates. }
\centering
\begin{tabular}{cccc}
 \hline 
$n_{z}$ & Circuit depth & Circuit width & $N_{z}$ \\
\hline
2 & 60   & 3   & 4 \\
3 & 182  & 5   & 8 \\
4 & 376  & 7   & 16 \\
5 & 642  & 9   & 32 \\
6 & 980  & 11  & 64 \\
7 & 1390 & 13  & 128 \\
 \hline  
\end{tabular} 
\label{table:ress_req}
\end{table}

It may be possible to implement our numerical scheme on quantum computing devices based on superconducting circuits. As mentioned earlier, this kind of quantum computer has been used successfully to perform digital quantum calculations \cite{barends2016digitized,barends2015digital,kelly2015state}. Using \texttt{Quipper}, our algorithm can be decomposed into a sequence of quantum logic gates, which can be implemented on superconducting circuits computers as a sequence of electric pulses. The gate decomposition of one time step for $n_{z}=3$ of the 1-D massive Dirac equation (see Table \ref{table:ress_req}) is given in Supplementary Material. The decomposition yields 72 \texttt{H}, 92 \texttt{S}, 94 \texttt{T} and 50 \texttt{CNOT} gates, for a total of 308 gates and a quantum register of 5 qubits. These operations can be carried on a superconducting circuits quantum computer. Assuming the number of operations for the initialization is low enough, a few time steps could be performed.

\section{Conclusion}
\label{sec:conclusion}

The analysis presented in this article have demonstrated that it is possible to solve the discrete Dirac equation efficiently on a quantum computer, including the initialization of the quantum register to an eigenstate of a static potential, under some conditions. Therefore, the technique presented in this article could be used to simulate important problems requiring a time-dependent solution of the Dirac equation such as pair production in Schwinger-like processes \cite{RevModPhys.84.1177} or the dynamics of charge carriers in graphene \cite{novoselov2005two}, for example.  

An explicit gate decomposition was carried out to evaluate the resource requirements and the feasibility of simulating relativistic quantum dynamics with actual quantum devices. It was demonstrated that the coherence time of existing quantum computers may allow for the evaluation of a few time iterations. Therefore, proof-of-principle calculations may be performed in the short term but a physically relevant calculation clearly necessitates much improvement in both the coherence time and in the number of qubits of quantum registers.

In this work, general electromagnetic fields have not been considered. General potentials entails diagonal operations which are similar to those found in Figs. \ref{fig:uniformctrl} - \ref{fig:init_wf} of Appendix \ref{sec:gen_init}. These necessitates an exponential number of gates, killing the performance of the quantum algorithm. Rather, we examined the possibility of having electromagnetic field with a vanishing magnetic field. As argued previously, for a large class of scalar potential, these can be included efficiently. In principle, a magnetic field could also be included by making the vector potential $\mathbf{A}$ space-dependent. The  quantum circuit that implements this effect is shown in Appendix \ref{sec:magnetic}. A general magnetic field requires an exponential number of gates, but as for the scalar potential, there may be special cases where it could be implemented in a logarithmic number of gates. Then, the translation invariance requirement can be relaxed completely. A thorough study of potentials implementable with a logarithmic number of gates is out of scope of this article, but would deserve more investigation.    

A possible extension of this work is for the Dirac equation in curved space time. It has been demonstrated in Ref. \cite{PhysRevA.88.042301,QW3} that the continuum limit of a certain class of space-dependent quantum walks reduces to the massless Dirac equation in a gravitational field. Then, it is plausible that the techniques presented in this article, also based on the analogy between quantum walks and the Dirac equation, could be applied to the gravitational case.  

Finally, it would be interesting to look at the possibility of simulating the Dirac equation by combining both analog and digital approaches, as proposed in Ref. \cite{arrazola2016digital}. Trapped ion quantum computers seems particularly suitable for this task, given that their quantum dynamics is analogous to the free part of the Dirac equation \cite{PhysRevLett.98.253005,gerritsma2010quantum}. It may be possible to take advantage of this, while keeping some aspects of the scheme given in this article, to obtain a more efficient algorithm. This is presently under investigation.

\begin{acknowledgments}
The authors would like to thank A.D. Bandrauk, E. Lorin, S. Succi and S. Palpacelli for many discussions relating to the numerical solution of the Dirac equation. Also, the authors acknowledge some important comments made by R. Somma on the initialization method and P. Selinger for some help with the code \texttt{Quipper}. Finally, we thank T. Farrelly for his useful comments on the fermion doubling problem.  
\end{acknowledgments}

\appendix

\section{Higher order schemes}
\label{sec:high_order}

An operator splitting scheme with a third order accuracy is given by \cite{suzuki1993general,FillionGourdeau20121403,Lorin2011190}:
%
\begin{align}
\psi(t_{n+1})
&= e^{-i\frac{\Delta t}{2}  \mathcal{T}} e^{-i\frac{\Delta t}{2} \hat{H}_{x} } e^{-i\frac{\Delta t}{2} \hat{H}_{y} } e^{-i\frac{\Delta t}{2} \hat{H}_{z} } e^{-i\frac{\Delta t}{2} \hat{H}_{m} }
\nonumber \\
& \times
 e^{-i\frac{\Delta t}{2} \hat{H}_{V}(t_{n}) }
e^{-i\frac{\Delta t}{2} \hat{H}_{\mathbf{A}}(t_{n}) }
 e^{-i\frac{\Delta t}{2} \hat{H}_{\mathbf{A}}(t_{n}) } e^{-i\frac{\Delta t}{2} \hat{H}_{V}(t_{n}) } 
\nonumber \\
& \times
e^{-i\frac{\Delta t}{2} \hat{H}_{m} }  e^{-i\frac{\Delta t}{2} \hat{H}_{z} } e^{-i\frac{\Delta t}{2} \hat{H}_{y} }e^{-i\frac{\Delta t}{2} \hat{H}_{x} }  e^{-i\frac{\Delta t}{2}  \mathcal{T}} \psi(t_{n}) 
\nonumber \\
& \times 
+O(\Delta t^{3}) ,\nonumber \\
 &= Q_{x}\left(\frac{\Delta t}{2}\right) Q_{y}\left(\frac{\Delta t}{2}\right) Q_{z}\left(\frac{\Delta t}{2}\right)  Q_{m}\left(\frac{\Delta t}{2}\right)
 \nonumber \\
 & \times
Q_{V}\left(t_{n}+\frac{\Delta t}{2},\frac{\Delta t}{2}\right) 
  Q_{\mathbf{A}}\left(t_{n}+\frac{\Delta t}{2},\Delta t\right) 
  \nonumber \\
  & \times
  Q_{V}\left(t_{n}+\frac{\Delta t}{2},\frac{\Delta t}{2}\right) 
  Q_{m}\left(\frac{\Delta t}{2}\right) 
 Q_{z}\left(\frac{\Delta t}{2}\right)
 \nonumber \\
 & \times
  Q_{y}\left(\frac{\Delta t}{2}\right) Q_{x}\left(\frac{\Delta t}{2}\right) \psi(t_{n})  +O(\Delta t^{3}).
\end{align}
%
Using the same strategy as for the second order scheme and the same type of discretization yields
%
\begin{align}
\label{eq:QLB_exp_order2_sp}
\psi_{\ell} (t_{n+1}) &= 
\left[S_{x}T_{x}\left(\frac{N^{*}\ell}{2} \right)S_{x}^{-1}\right]
\left[S_{y}T_{y}\left(\frac{N^{*}\ell}{2} \right)S_{y}^{-1}\right]
\nonumber \\
& \times
\left[S_{z}T_{z}\left(\frac{N^{*}\ell}{2} \right)S_{z}^{-1}\right]
Q_{m}\left(\frac{\Delta t}{2} \right)
\nonumber \\
& \times
Q_{V}\left(t_{n}+\frac{\Delta t}{2},\frac{\Delta t}{2}\right) 
Q_{\mathbf{A}}\left(t_{n}+\frac{\Delta t}{2},\Delta t\right)
\nonumber \\
& \times
Q_{V}\left(t_{n}+\frac{\Delta t}{2},\frac{\Delta t}{2}\right) 
Q_{m}\left(\frac{\Delta t}{2} \right)
 \nonumber \\
& \times
\left[S_{z}T_{z}\left(\frac{N^{*}\ell}{2} \right)S_{z}^{-1}\right]
\left[S_{y}T_{y}\left(\frac{N^{*}\ell}{2} \right)S_{y}^{-1}\right]
\nonumber \\
& \times
\left[S_{x}T_{x}\left(\frac{N^{*}\ell}{2} \right)S_{x}^{-1}\right] 
\psi_{\ell} (t_{n}) .
\end{align}

An $m$'th order splitting can be obtained from the $m-1$'th order splitting using Suzuki's iterative scheme \cite{Suzuki1990319}. The latter states that the $m$'th order approximant $F_{m}(\Delta t)$, which yields an error as $O(\Delta t^{m+1})$, is given in terms of the $m-1$'th order approximant as
\begin{align}
F_{m}(\Delta t) = F_{m-1}(p_{1}\Delta t) \cdots F_{m-1}(p_{r}\Delta t),
\end{align}
where $r \in \mathbb{N}^{+}$ while the parameters $p_{1},\cdots,p_{r} \in \mathbb{C}$ are constrained by the following equations:
\begin{align}
\label{eq:suzuki_condition}
\sum_{i=1}^{r} p_{i} = 1 \;\; , \;\; \sum_{i=1}^{r} p^{m}_{i} = 0.
\end{align}
The value of $r$ is chosen arbitrarily, but in practice, it is important to use the smallest value of $r$ as possible to reduce the number of operations. For a given $r$, the solution of Eq. \eqref{eq:suzuki_condition} is not necessarily unique: the best choice then is essentially a matter of convenience. 

\begin{table}
\caption{Possible rational splittings schemes for $m=3$ with an accuracy $O(\Delta t^{4})$. }
\centering
\begin{tabular}{cccccccccc}
 \hline 
 $r$ & $\tilde{p}_{1}$ & $\tilde{p}_{2}$ & $\tilde{p}_{3}$ & $\tilde{p}_{4}$ & $\tilde{p}_{5}$& $\tilde{p}_{6}$& $\tilde{p}_{7}$& $\tilde{p}_{8}$& $\tilde{p}_{9}$\\
 \hline  
7& 6& 6& 6& 3& 3& 3& -2 \\
8& 6& 4& 4& 4& 3& 3& -2& -12 \\
9 & 6& 6& 6& 6& 6& 6& 6& 6& -3 \\
9&6 &6& 6& 3& 3& 3& 2& -2& -2  \\
9&6 &6& 6& 3& 3& 3& 3& -2& -3 \\
9&12& 6& 6& 6& 3& 3& 3& -2& -12  \\
 \hline  
\end{tabular} 
\label{table:splitting_fourth}
\end{table}

To obtain a splitting where the parameters $p_{i}$ are rational numbers and where every $p_{i}$ is a multiple of the smallest one, it is convenient to define $\tilde{p}_{i} = 1/p_{i}$ for $i=1,\cdots r$. Then, the splitting we are looking for should obey the following conditions:
\begin{align}
\begin{cases}
(\tilde{p}_{i})_{i=1,\cdots, r} \in \mathbb{N} , \\
(p_{i} = n_{i}p_{j})_{i=1,\cdots, j-1,j+1,\cdots , r}, n_{i} \in \mathbb{N} \\
\quad \quad \quad \quad \quad \quad \quad \mbox{for} \;\; p_{j} \leq (p_{i})_{i=1,\cdots, j-1,j+1,\cdots , r} ,\\
\sum_{i=1}^{r} \frac{\prod_{j=1}^{r} \tilde{p}_{j}}{ \tilde{p}_{i}} = \prod_{i=1}^{r} \tilde{p}_{i} ,\\
\sum_{i=1}^{r} \frac{\prod_{j=1}^{r} \tilde{p}^{m}_{j}}{ \tilde{p}^{m}_{i}} = 0.
\end{cases}
\end{align}  
This system of equation is challenging to solve for large $m$ and $r$. For $m=3$, solutions shown in Table \ref{table:splitting_fourth} can be found by a systematic searching algorithm. There is no solution for $r<7$.

\section{Quantum circuit for the inclusion of a magnetic field}
\label{sec:magnetic}

\begin{figure*}
\includegraphics[scale=1.0]{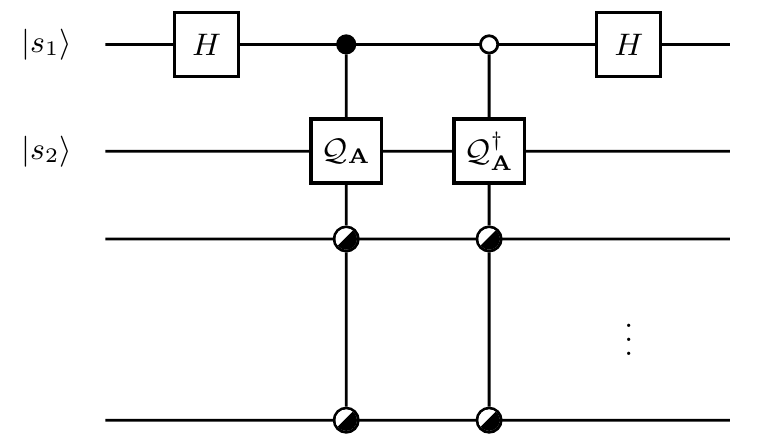}
\caption{Quantum circuit for the inclusion of a magnetic field. }
\label{fig:magnetic}
\end{figure*}

When a magnetic field is included, the vector potential $\mathbf{A}$ depends on both time and space. In this case, the decomposition in Eq. \eqref{eq:V_euler} still holds but then, the rotation angles in Eqs. \eqref{eq:rot_angle_delta} to \eqref{eq:rot_angle_theta} depends on space as $\delta_{i,j,k},\xi_{i,j,k}$ and $\theta_{i,j,k}$. The space dependence can be introduced by using uniformly controlled gates, as displayed in Fig. \ref{fig:magnetic}. Such circuit are not efficient for general space dependence because they require $N_{x}N_{y}N_{z}$ multi-controlled gates. However, it may be possible to find special cases where the vector potential can be implemented in $\mathrm{poly}(n_{x},n_{y},n_{z})$. 

\section{General initial states}
\label{sec:gen_init}

\begin{figure*}
\includegraphics[scale=1.0]{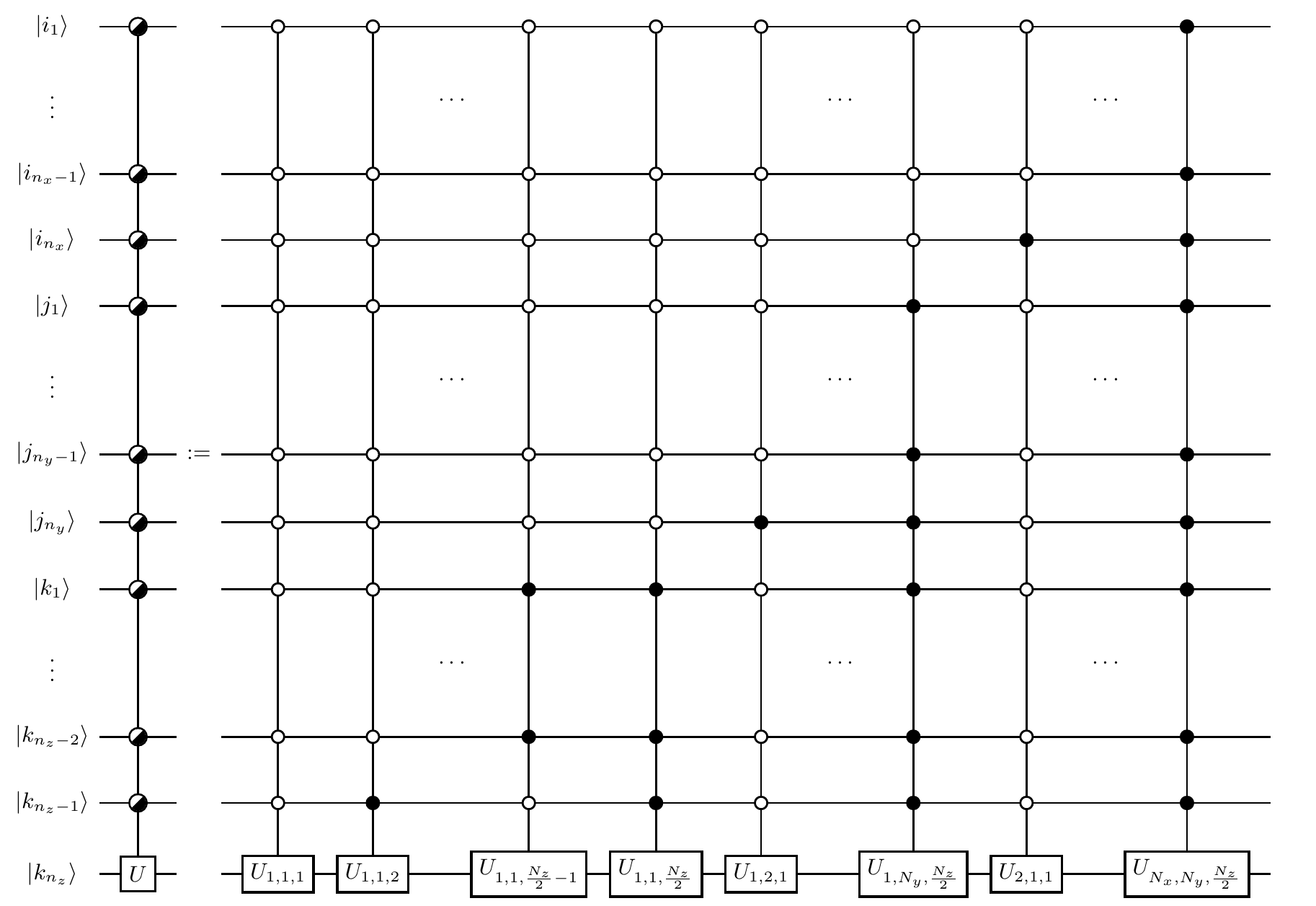}
\caption{Definition and circuit diagram for the grouped uniformly controlled gate for $k+1$ qubits \cite{PhysRevA.71.052330}. The half black-white controls on the left-hand-side imply a sum on all black and white controls with a different operator $U_{i,j,k}$.}
\label{fig:uniformctrl}
\end{figure*}

\begin{figure}
\includegraphics[scale=1.0]{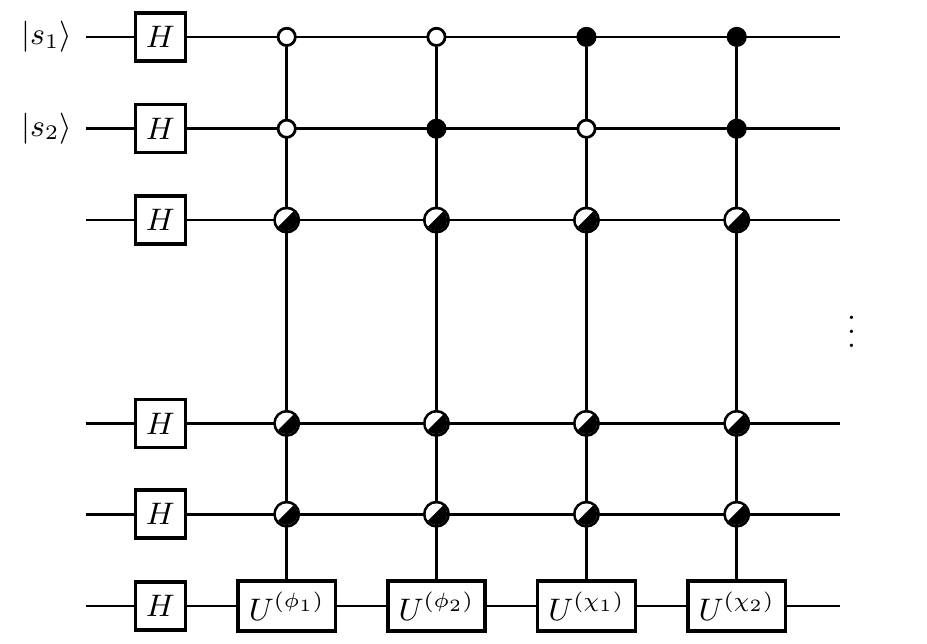}
\caption{Circuit diagram for the initialization of the wave function.}
\label{fig:init_wf}
\end{figure}

The initialization for an initial state corresponds to the following mapping (here given for one arbitrary spinor component $\psi_{S}$):
\begin{widetext}
\begin{align}
\label{eq:map_init}
\frac{1}{2^{\frac{n}{2}}}
\begin{pmatrix}
1\\1\\ \vdots \\1
\end{pmatrix}
\rightarrow
\begin{pmatrix}
\psi_{S,\mathrm{init}}(\tilde{\mathbf{x}}_{0,0,0})\\
\psi_{S,\mathrm{init}}(\tilde{\mathbf{x}}_{0,0,1})\\ 
\vdots \\
\psi_{S,\mathrm{init}}(\tilde{\mathbf{x}}_{N_{x},N_{y},N_{z}})
\end{pmatrix} =
\begin{pmatrix}
|\psi_{S,\mathrm{init}}(\tilde{\mathbf{x}}_{0,0,0})|e^{i\varphi_{S,0,0,0}}\\
|\psi_{S,\mathrm{init}}(\tilde{\mathbf{x}}_{0,0,1})|e^{i\varphi_{S,0,0,1}}\\ 
\vdots \\
|\psi_{S,\mathrm{init}}(\tilde{\mathbf{x}}_{N_{x},N_{y},N_{z}})|e^{i\varphi_{S,N_{x},N_{y},N_{z}}}
\end{pmatrix},
\end{align}
\end{widetext}
where $\varphi_{S,i,j,k}$ is the phase of the wave component $\psi_{S,\mathrm{init}}(\tilde{\mathbf{x}}_{i,j,k})$. 
Assuming that the quantum register is initialized in the state $|00\cdots 0\rangle$, the left part of Eq. \eqref{eq:map_init} can be obtained via a Hadamard transform. The mapping \eqref{eq:map_init} is a diagonal operation that can be realized by a sequence of uniformly-controlled quantum gates (UCQG) \cite{PhysRevA.71.052330}: this class of gates is defined in Fig. \ref{fig:uniformctrl} and they consist in the set of all possible multi-qubits controlled gates. Then, the initialization proceeds by using one gate per spinor component, as displayed in Fig. \ref{fig:init_wf}. 

The operators $U_{i,j,k}^{(\psi_{S})}$ appearing in the gate definition are related to the value of the wave function. First, they are 1-qubit operation and thus, can be decomposed as a sequence of rotation operators as follows:
\begin{align}
U_{i,j,k}^{(\psi_{S})} = e^{i\gamma_{S,i,j,k}}R_{z}(\delta_{S,i,j,k})R_{y}(\theta_{S,i,j,k}),
\end{align} 
where the parameters $\gamma_{S,i,j,k},\delta_{S,i,j,k},\theta_{S,i,j,k} \in [0,2\pi]$ characterize the unitary operation. These parameters are related to the initial wave function as
\begin{align}
\theta_{S,i,j,k} &= 2 \arccos \left( \cfrac{|\psi_{S,\mathrm{init}}(\tilde{\mathbf{x}}_{i,j,2k-1})| + |\psi_{S,\mathrm{init}}(\tilde{\mathbf{x}}_{i,j,2k})|}{2} \right),\nonumber \\ \\
\gamma_{S,i,j,k} &= \varphi_{S,i,j,2k} + \varphi_{S,i,j,2k-1},\\
\delta_{S,i,j,k} &= \varphi_{S,i,j,2k} - \varphi_{S,i,j,2k-1}.
\end{align}
This unitary operation initializes the wave function at two points simultaneously, hence the limit of the index $k \in [1,N_{z}/2]$ in the circuit diagram of Fig. \ref{fig:uniformctrl}.

The number of gates required to initialize the wave function on the quantum register can be evaluated by analysing the UCQGs. It has been demonstrated that the complexity of an arbitrary $n$-qubit UCQG is $O(4^{n})$ \cite{PhysRevA.71.052330}. This can be improved further to $O(2^{n})$ \cite{PhysRevA.83.032302}. Applying these results to the Dirac equation, we obtain that the number of quantum gates should scale like $O(2^{n_{x}}2^{n_{y}}2^{n_{z}})$. Therefore, this part of the algorithm is not efficient because it is not a polynomial of $n_{x},n_{y},n_{z}$ for the initialization of a general space-time dependent wave function.

\section{Quantum Feit-Fleck method}
\label{sec:init}

In this appendix, the quantum Feit-Fleck method given in Ref. \cite{ffg.feitfleck} is reviewed and applied to the Dirac equation. The Feit-Fleck spectral method has been originally developed to evaluate eigenenergies and eigenstates of the Schr\"{o}dinger equation in a static potential \cite{Feit1982412}. This technique was then applied to the Dirac equation to calculate eigenfunctions of hydrogen-like atoms \cite{Mocken2004558,Mocken2008868,PhysRevA.83.063414,Bauke2011}. It is well suited for Dirac operators because it does not require the spectrum to be bounded from below, in contrast with variational methods.

\begin{figure}
\includegraphics[scale=1.0]{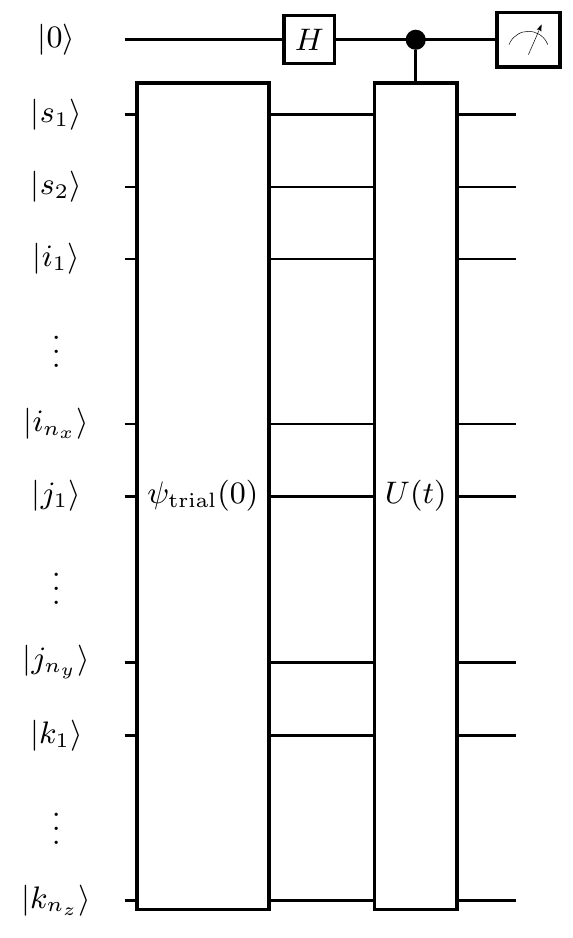}
\caption{Circuit diagram for the computation of the autocorrelation function. The quantity $\langle (\sigma_{x}+i\sigma_{y})\otimes \mathbb{I} \rangle$ is measured on the ancilla qubit.}
\label{fig:autocorr}
\end{figure}

An accurate approximation of the eigenenergy of the desired eigenstate is expected for the filtering phase presented below. In some cases, these eigenenergies can be estimated from analytical methods or classical computations. However, there exist techniques to perform this task efficiently on a quantum computer. The starting point is the autocorrelation function $C(t)$, given by \cite{Feit1982412}
\begin{align}
\label{eq:auto_corr}
C(t) &=  \int d^{3}\mathbf{x} \psi^{*}(0,\mathbf{x})\psi(t,\mathbf{x}) \\
C(E) &= \frac{1}{t_{f}}\int_{0}^{t_{f}} dt w(t) e^{iEt}C(t).
\end{align}  
The eigenenergies appear as sharp peaks in the spectral density $C(E)$. This autocorrelation can be computed semi-classically \cite{PhysRevLett.76.3228,ffg.feitfleck,PhysRevLett.81.5672,PhysRevA.65.042323}. First, an ancilla qubit is added in the state $|0\rangle$. Applying a Hadamard gate on this ancilla qubit and initializing some arbitrary trial state, the quantum register will be in the state
\begin{align}
&\frac{1}{\sqrt{2}} \left[|0\rangle + |1\rangle \right]\otimes |\psi_{\mathrm{trial}}(0)\rangle , \\
\label{eq:register}
&\mapsto \frac{1}{\sqrt{2}} \left[|0\rangle \otimes |\psi_{\mathrm{trial}}(0)\rangle + |1\rangle \otimes |\psi_{\mathrm{trial}}(t)\rangle \right].
\end{align}
The mapping in Eq. \eqref{eq:register} is obtained by applying a controlled evolution operator that evolves the trial state to some final time $t$. Then, it can be demonstrated that performing the following measurement on the ancilla qubit yields the autocorrelation function: 
\begin{align}
\langle (\sigma_{x}+i\sigma_{y})\otimes \mathbb{I} \rangle &= \langle \psi_{\mathrm{trial}}(0) |\psi_{\mathrm{trial}}(t)\rangle , \\
\label{eq:quantum_autocorr}
&= \sum_{S=1}^{4} \sum_{i=1}^{N_{x}} \sum_{j=1}^{N_{y}} \sum_{k=1}^{N_{z}} \alpha^{*}_{S,i,j,k}(0)\alpha_{S,i,j,k}(t), \nonumber \\
\end{align}
where $\alpha_{S,i,j,k}(t)$ are the coefficients of the register that store the wave function, as in Eq. \eqref{eq:coeff_regis}. Eq. \eqref{eq:quantum_autocorr} is a discretized version of the autocorrelation function in Eq. \eqref{eq:auto_corr}. The resulting circuit diagram is displayed in Fig. \ref{fig:autocorr}. 

Once the autocorrelation has been evaluated, the Fourier transform has to be performed classically. Then, the eigenenergies can be read off the spectral density $C(E)$.

Here, the final time $t_{f}$, where $t\in [0,t_{f}]$, is important because it determines the resolution of the spectral method as $\Delta E = \pi/t_{f}$, where $\Delta E$ is the energy resolution of the method. Therefore, obtaining a precise value of the eigenenergy requires a large simulation time. Moreover, the energy range that can be considered is governed by the time step as $[-\pi/\Delta t,\pi/\Delta t]$.  



Once the eigenenergy is known, it is possible to use the following equation to filter an arbitrary trial state \cite{Feit1982412}:
\begin{align}
\label{eq:eigen_state_ff}
\psi_{E}(\mathbf{x}) = \frac{1}{t_{f}} \int_{0}^{t_{f}} dt \psi_{\mathrm{trial}}(t,\mathbf{x}) w(t)e^{iEt},
\end{align}
where $\psi_{E}(\mathbf{x})$ is the wanted eigenstate, $E$ is the energy of the eigenstate, $t_{f}$ is is final time of the calculation, $\psi_{\mathrm{trial}}(t,\mathbf{x})$ is an arbitrary trial function and $w(t)$ is a window function. One convenient choice for the window function is the Hann function but other choices are available \cite{heinzel2002spectrum}.



The filtering can be implemented on a quantum computer by supplementing the quantum register with an additional qubit $|c\rangle$. Then, 
Eq. \eqref{eq:eigen_state_ff} is approximated by a quadrature formula of the form
\begin{align}
\label{eq:eigen_state_quad}
\psi_{E}(\mathbf{x}) &\approx   \sum_{k=0}^{N_{t}} B_{k} \psi_{\mathrm{trial}}(t_{k},\mathbf{x}) ,\\
B_{k} &:= \Delta t  a_{k} w(t_{k})e^{iEt_{k}},
%
\end{align}
where $N_{t}$ is the number of timestep, $t_{k} = k\Delta t$ is the time where the integrand is evaluated and $(a_{k})_{k=0,\cdots,N_{t}}$ are coefficients required by the quadrature rule \footnote{For example, for the trapezoidal rule, we have $a_{0} = a_{N_{t}} = 1/2$ and $(a_{k})_{k=1,\cdots,N_{t}-1}=1$.}. 

\begin{figure*}
\includegraphics[scale=1.0]{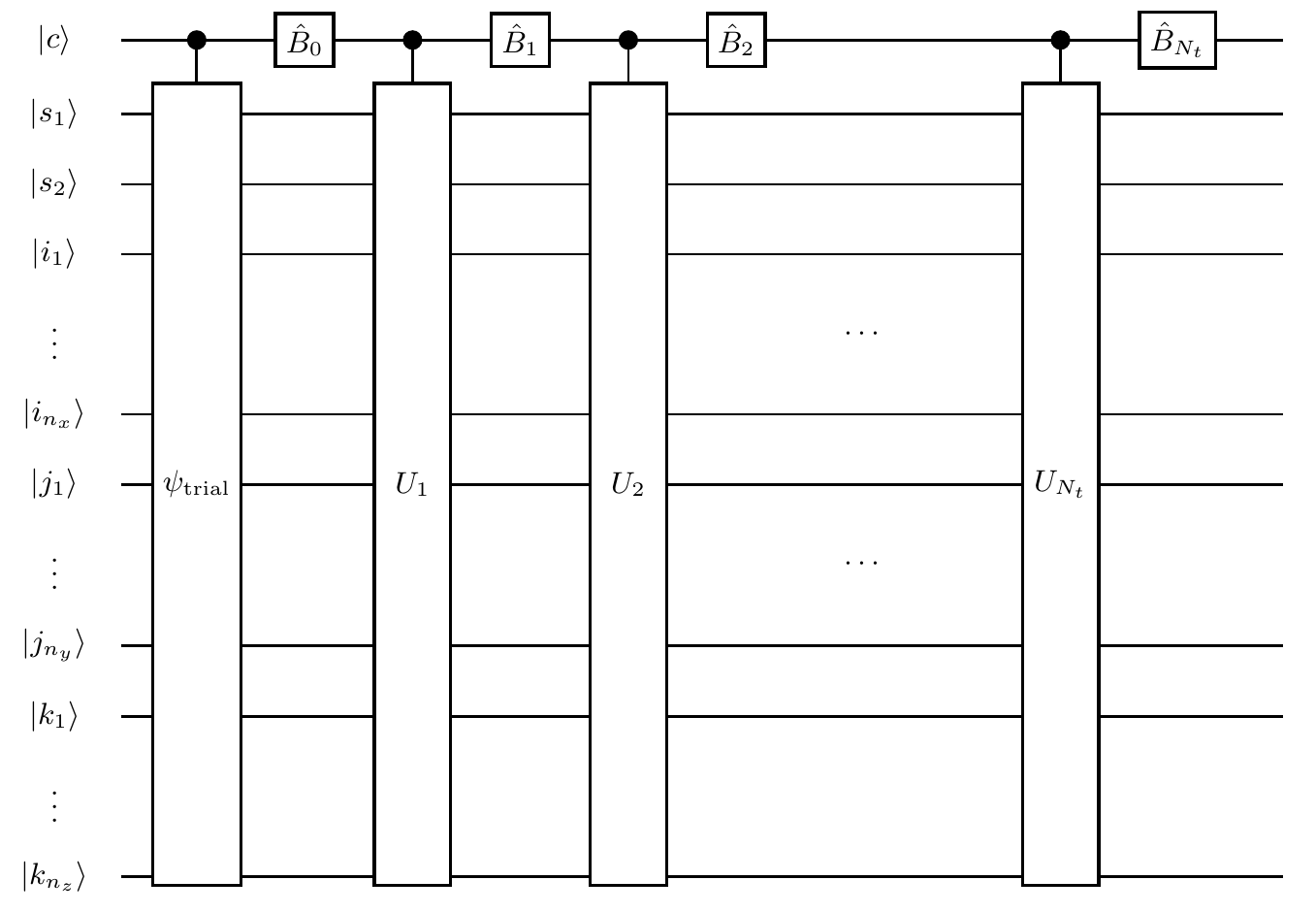}
\caption{Circuit diagram for the quantum implementation of the Feit-Fleck method. The gate $U_{i} := U((i-1)\Delta t,i\Delta t)$ advances the solution by $\Delta t$, its decomposition into quantum gates is given in Fig. \ref{fig:timestep}. The gate $B_{i}$ is defined in Eq. \eqref{eq:gate_B}.}
\label{fig:init_ff}
\end{figure*}

The result of the partial sum can be stored in the register by applying a non-unitary operator at every time step defined by $\hat{B}_{i} \otimes \mathbb{I}_{2} \cdots \otimes \mathbb{I}_{2}$ where $\hat{B}_{i}$ is a two-by-two matrix given by \cite{ffg.feitfleck}
\begin{align}
\label{eq:gate_B}
\hat{B}_{i} := \frac{1}{ \sqrt{1+ \frac{|B_{i}|^{2}}{2} + |B_{i}|\sqrt{1+\frac{|B_{i}|^{2}}{4}}}}
\begin{bmatrix}
1 & 0 \\
B_{i} & 1
\end{bmatrix}.
\end{align} 
The resulting quantum circuit is displayed in Fig. \ref{fig:init_ff}. 

\begin{figure}
\includegraphics[scale=1.0]{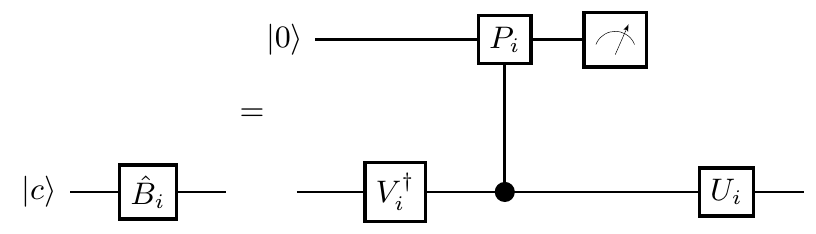}
\caption{Circuit diagram for the implementation of the nonunitary operation. The upper qubit is an ancilla qubit prepared in the state $|0\rangle$. The measurement operator implements the projective measurement $|0\rangle \langle 0|$. If the measurement yields the state $|1\rangle$, the calculation has to be redone from the beginning.}
\label{fig:nonunitary}
\end{figure}

It is possible to implement non-unitary operations on a quantum computer by using non-deterministic algorithms \cite{Terashima2005,mezzacapo2015quantum,Blass2015,Childs:2012:HSU:2481569.2481570,PhysRevLett.114.090502}. Following the technique described in \cite{Terashima2005}, the first step is to find the singular value decomposition of the matrix $\hat{B}$. It is given by
\begin{align}
\label{eq:SVD}
\hat{B}_{i} &= U_{i} \Sigma_{i} V^{\dagger}_{i}. 
\end{align}
The matrices $U_{i}, V^{\dagger}_{i}$ are unitary while $\Sigma_{i} = \mathrm{diag}(1,a_{i})$ is diagonal, where the singular value is given by
\begin{align}
a_{i} &= \sqrt{ \frac{1+ \frac{|B_{i}|^{2}}{2} - |B_{i}|\sqrt{1+\frac{|B_{i}|^{2}}{4}}}{1+ \frac{|B_{i}|^{2}}{2} + |B_{i}|\sqrt{1+\frac{|B_{i}|^{2}}{4}}}} ,
\end{align}
where $a_{i} \leq 1$, in accordance with the exact realization theorem \cite{Blass2015}. Then, the operator $\Sigma_{i}$ can be literally realized with one ancilla initialized in the state $|0\rangle$, a unitary transformation and a projective measurement \cite{ffg.feitfleck}. The corresponding quantum circuit is displayed in Fig. \ref{fig:nonunitary}, where the controlled unitary operator is given by 
\begin{align}
\mbox{c-}P_{i} = 
\begin{bmatrix}
1 & 0 & 0 & 0 \\
0 & a_{i} & 0 & \sqrt{1-a_{i}^{2}} \\
0 & 0 & 1 & 0 \\
0 & -\sqrt{1-a_{i}^{2}}& 0 & a_{i}
\end{bmatrix}.
\end{align}
The last step of the circuit is a projective measurement $|0\rangle \langle 0|$ on the ancilla qubit. A success occurs when the ancilla is measured in the state $|0\rangle$, which implies that the non-unitary operation has been implemented properly. The success probability of this projective measurement, after $N_{t}+1$ iterations is \cite{ffg.feitfleck}
\begin{align}
\label{eq:prob_success}
P_{\mathrm{success}}(N_{t}+1) &\geq  \ \frac{1}{e} \left[ 1 - \frac{1}{N_{t}}  \right] + O\left(\frac{1}{N_{t}^{2}} \right),
\end{align}
where $e \approx 2.7183$ is Euler's number (not the electric charge).

The number of operations required to initialize the wave function using this quantum implementation of the Feit-Fleck method scales like $\bar{N} = e^{2}N_{t}\mathrm{poly}(n_{x},n_{y},n_{z})/P$, where $P$ is the probability to be in the eigenstate after the filtering. As long as $P$ is not exponentially small, the initialization can be performed using a logarithmic number of gates. The performance is similar to the phase-estimation algorithm but requires less ancilla qubits. More details are given in Ref. \cite{ffg.feitfleck}.

\bibliographystyle{apsrev4-1}
\bibliography{bibliography}

\end{document}